\documentclass{article}
\usepackage[utf8]{inputenc}
\usepackage[T1]{fontenc}
\usepackage{arxiv}
\usepackage{titlesec}
\usepackage{booktabs}
\usepackage{array}
\usepackage{enumitem}
\usepackage{amssymb}
\usepackage{amsmath}
\usepackage{xcolor}
\usepackage{graphicx}
\usepackage{float}

\usepackage{caption}
\usepackage{subcaption}
\usepackage[numbers,sort&compress]{natbib}
\usepackage{hyperref}
\usepackage{indentfirst}
\usepackage{pgfplots}
\pgfplotsset{compat=1.18}
\usepgfplotslibrary{groupplots}
\hypersetup{
    colorlinks=true,
    linkcolor=blue,
    filecolor=magenta,
    urlcolor=blue,
    citecolor=blue,
    breaklinks=true
}
\usepackage{xurl}
\titleformat{\section}[block]{\Large\bfseries}{\thesection.}{0.5em}{}
\titleformat{\subsection}[block]{\normalsize\bfseries}{\thesubsection}{0.5em}{}

\title{\textbf{Design and Experimental Evaluation of a Low-Cost Immersion Cooling System}}

\author{
Pramukto Mandal$^{b}$, Bhavya Jain$^{b}$, M.U.~Kusheetha$^{a}$, Nikhil Meena$^{c}$,\\
\bfseries Ravindra Kumar$^{c}$, M.P.~Sabarish$^{c}$, Amit Verma$^{b}$, Gaurav Bhutani$^{*}$\\[8pt]
\small $^{a}$School of Mathematics and Statistical Sciences, IIT Mandi, India\\
\small $^{b}$School of Mechanical and Materials Engineering, IIT Mandi, India\\
\small $^{c}$School of Computing and Electrical Engineering, IIT Mandi, India\\
\small $^{*}$Corresponding author: gaurav@iitmandi.ac.in}

\date{}

\renewcommand{\headeright}{}
\renewcommand{\undertitle}{}

\begin{document}

\maketitle

\begin{abstract}
This paper presents a low-cost, open-source, single-phase immersion cooling testbed for studying liquid cooling of data center hardware. A 500~W titanium heater, standing in for server load, is submerged in a glass tank of mineral oil, a dielectric fluid. The heated oil is circulated by a magnetic drive pump through an air-cooled radiator. An ESP32 microcontroller logs three DS18B20 temperature sensors (heater surface, bulk fluid and ambient) and switches the pump, fan and heater through a relay module. The complete system costs INR~35,318 (about USD~380), built entirely from components available in India. At quasi-steady state under the full 500~W load, the heater surface reached 118.3~\textdegree{}C while the bulk oil stayed at 29.3~\textdegree{}C, only 8.0~\textdegree{}C above ambient. Thermal-resistance analysis shows that the heater-to-oil interface, limited by natural convection, accounts for about 92\% of the total resistance (178.0 vs 16.0~m\textdegree{}C/W). The external loop therefore has spare capacity. CAD files, firmware and data are openly available, so the setup can be reproduced and used to compare dielectric fluids, loads and flow arrangements.
\end{abstract}


\bigskip
\noindent\textbf{Specifications table}

\vskip 0.2cm
\begin{table}[H]
\centering
\sloppy
\begin{tabular}{>{\raggedright\arraybackslash}p{0.26\linewidth}p{0.68\linewidth}}\toprule

\textbf{Hardware name} & \textit{Custom Immersion Cooling System for Data Centers} \\\midrule

\textbf{Subject area} & \textbullet~\textit{Engineering and material science}\newline
\textbullet~\textit{Educational tools and open source alternatives to existing infrastructure}\\

\textbf{Hardware type} & \textbullet~\textit{Measuring physical properties and in-lab sensors}\newline
\textbullet~\textit{Mechanical engineering and materials science}\\

\textbf{Closest commercial analog} & \textit{This hardware is an open source alternative to commercial immersion systems. The closest analogs are manufactured by Submer, Green Revolution Cooling (GRC) \cite{ref1}, and Asperitas \cite{ref9,ref10}.} \\

\textbf{Open source license} & {\it CC-By Attribution 4.0 International} \\

\textbf{Cost of hardware} & \textit{INR 35318} \\

\textbf{Source file repository} & \url{https://doi.org/10.17605/OSF.IO/SWZV7}\\ \bottomrule

\end{tabular}
\end{table}

\section{Hardware in context}

These days, data centers are becoming necessary for almost all online services, 
including video streaming and cloud storage. Global data center electricity 
consumption is estimated to account for a significant and growing share of 
worldwide energy use, and this demand is expected to continue rising as cloud 
computing, artificial intelligence workloads, and high-performance computing expand \cite{ref5,ref12}. The amount of heat generated continues to increase as the 
number of data centers and the density of compute hardware within them increases 
\cite{ref12}. A fundamental challenge has therefore arisen in cooling all of that 
equipment while maintaining operational efficiency and reducing energy consumption 
\cite{ref4,ref8}.

\begin{figure}[!htbp]
    \centering
    
    \begin{subfigure}[b]{0.48\linewidth}
        \centering
        \includegraphics[width=\linewidth]{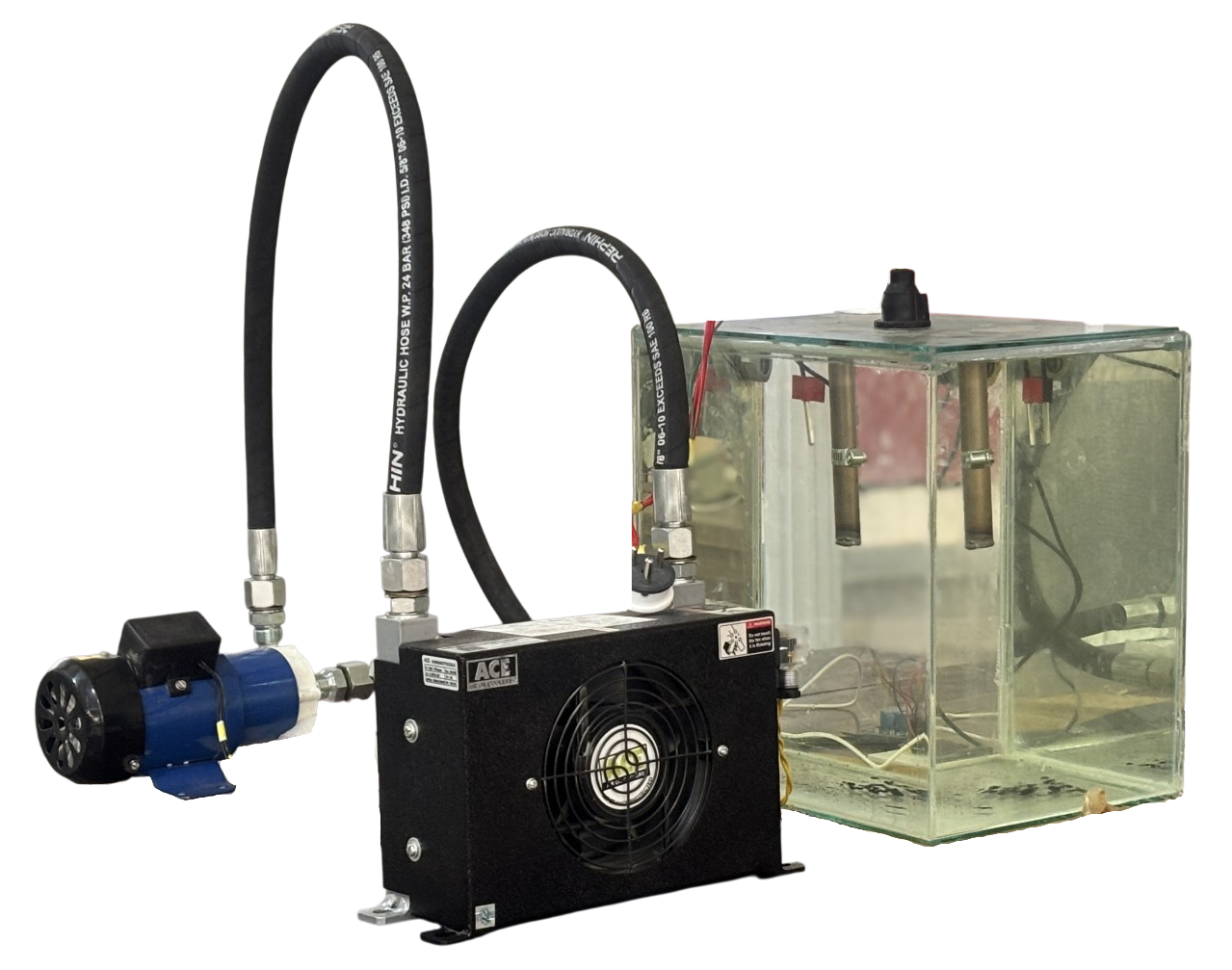}
        \caption{Photo of the setup}
        \label{fig:setup_photo}
    \end{subfigure}
    \hfill
    \begin{subfigure}[b]{0.48\linewidth}
        \centering
        \includegraphics[width=\linewidth]{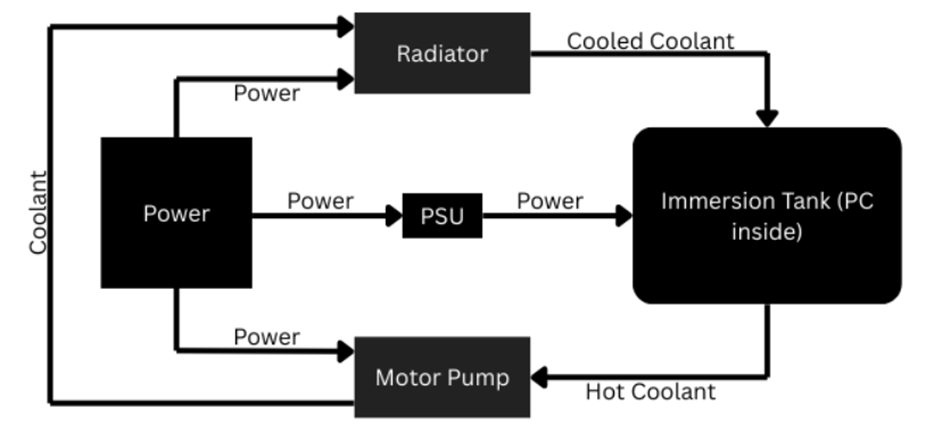}
        \caption{Block diagram}
        \label{fig:block_diagram}
    \end{subfigure}
    
    \caption{Low-cost immersion cooling system: (a) photo of the experimental setup and (b) block diagram.}
    \label{fig:setup}
\end{figure}

Most data centers rely on traditional air-conditioned cooling even today, which 
uses fans, computer room air conditioning (CRAC) units, and precision air handlers 
to remove heat from the servers \cite{ref12}. This approach typically involves 
carefully managed hot-aisle and cold-aisle arrangements to direct airflow across 
server racks. While this method is functional and well understood, it is not a 
thermally efficient solution: air has a relatively low convective heat transfer coefficient compared to liquid coolants \cite{bergman2011}, and maintaining the pressure and flow of conditioned air across 
large server halls requires significant infrastructure. As a result, a large 
fraction of the total power consumed by a data center goes into cooling alone \cite{ref5,ref8,ref12}. A common metric for quantifying this overhead is the Power 
Usage Effectiveness (PUE) ratio, defined as the total facility energy divided by 
the energy consumed by IT equipment alone \cite{iso30134}. The industry-average PUE has stayed at about 1.55--1.6 for several years \cite{ref12,uptime2024}, meaning that for every watt consumed by IT equipment, more than half a watt is spent on cooling and other infrastructure. This 
makes the system expensive to operate and places a disproportionate environmental 
burden on data center infrastructure \cite{ref12}. Heat pipes are also widely used 
as passive heat-transfer elements in compact heat exchangers and in electronics 
cooling applications, offering high effective thermal conductance with no moving 
parts \cite{ref2}, but in electronics cooling they are typically employed as localized solutions within individual components rather than as a system-level cooling strategy.

A more effective alternative that has gained significant traction in recent years 
is immersion cooling \cite{ref3,ref12}. This method involves submerging 
electronic components directly in a dielectric fluid --- one that does not conduct 
electricity and thus poses no risk of short-circuiting the hardware. Deionized water, mineral oil, fluorocarbon-based fluids and synthetic oils have all been used as immersion coolants \cite{ref3}. Mineral oil is the most commonly used fluid in single-phase systems \cite{ref4,ref12}; it is low-cost, readily available and compatible with standard electronic components \cite{ref8}. Engineered fluorocarbon fluids such as FC-40 have also been studied \cite{ref4}; FC-40 has a far lower kinematic viscosity than mineral oil (about 1.5~cSt at 40~\textdegree{}C \cite{ref4}, versus 9.9~cSt for the mineral oil used here \cite{ref13}), but fluorocarbon coolants typically cost an order of magnitude more than hydrocarbon oils \cite{ref1}. In all cases, the principle is the same: direct liquid contact with heat-generating surfaces transfers heat far more effectively than air cooling \cite{ref3,ref7}.

Immersion cooling systems can be broadly categorized into single-phase and 
two-phase variants \cite{ref1}. In single-phase immersion cooling, the dielectric 
fluid remains in its liquid state throughout the cooling cycle and heat is removed 
by circulating the warmed fluid through an external heat exchanger \cite{ref3,ref5,ref6,ref8}. 
In two-phase systems, the fluid is allowed to boil at the component surface, and 
the latent heat of vaporization is exploited for higher heat removal rates; the resulting vapor is then condensed back to liquid and returned to the tank \cite{ref1}. While two-phase systems are better suited to ultra-high heat fluxes \cite{ref4}, single-phase systems are simpler to implement, require no condenser infrastructure, 
and are fully compatible with conventional mineral oil, making them the more 
accessible choice for low-cost and experimental implementations \cite{ref1,ref4}.

Although commercial immersion cooling systems are more costly and primarily 
intended for industrial-scale applications, large companies have already 
demonstrated significant reductions in cooling energy consumption using this 
approach, with PUE values approaching 1.0 in some deployments \cite{ref1,ref12}. 
Commercial immersion cooling solutions are being developed by multiple vendors for both servers and high-performance PCs, including Submer, Green Revolution 
Cooling (GRC), Asperitas, and Thermaltake \cite{ref1,ref9,ref11}. Despite this momentum, 
access to immersion cooling as a hands-on experimental platform 
remains limited: commercial systems are expensive, closed-source, and designed for 
production deployment rather than pedagogical exploration.

The primary goal of our project is to address this gap by creating a manageable 
and low-cost version of a single-phase immersion cooling system, such that it can 
be readily constructed and demonstrated in a lab. As shown in Fig.~\ref{fig:setup}, the system consists 
of a glass tank filled with mineral oil as the dielectric fluid, a 500~W electrical heating element that replicates the thermal output of server hardware, a magnetic drive pump for fluid circulation, an 
air-cooled oil radiator for heat rejection, and three DS18B20 digital temperature 
sensors for real-time thermal monitoring. To efficiently demonstrate the cooling 
behavior and log the temperatures being measured, an ESP32 microcontroller was used
as the central control and data acquisition unit. A compact relay module is 
integrated into the system to control the 230~V AC components safely, with 
low-voltage control signals from the ESP32 enabling convenient and electrically 
isolated switching. This design promotes easily accessible research in sustainable 
cooling techniques and requires no specialized procurement --- every component in 
the bill of materials is locally available within India at a total hardware cost 
of INR~35,318 ($\approx$USD~380).

The system gives users the freedom to experiment with various dielectric fluids 
and under varying thermal loads. It is designed to be modular, allowing 
straightforward substitution of the pump, radiator, tank, or sensing elements for 
comparative experiments. 
This design provides a practical and low-cost platform for experimental investigation of immersion cooling concepts and for future development of improved thermal management strategies for data center applications.

\section{Hardware description}

\subsection{Immersion tank}

\begin{figure}[!htbp]
    \centering
    \begin{subfigure}[b]{0.30\linewidth}
        \centering
        \includegraphics[width=\linewidth]{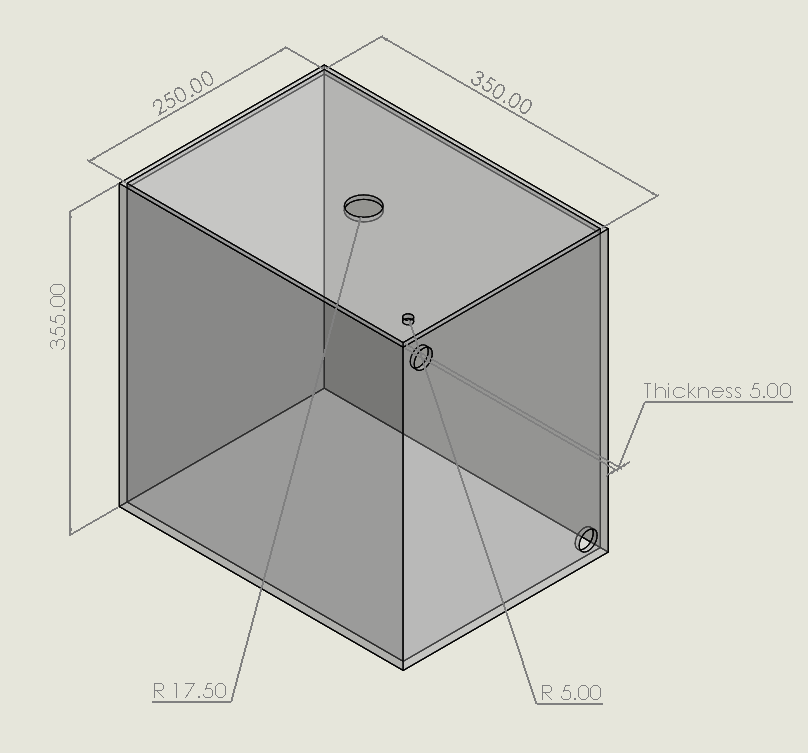}
        \caption{Immersion tank (CAD, mm).}
    \end{subfigure}
    \hfill
    \begin{subfigure}[b]{0.22\linewidth}
        \centering
        \includegraphics[width=\linewidth]{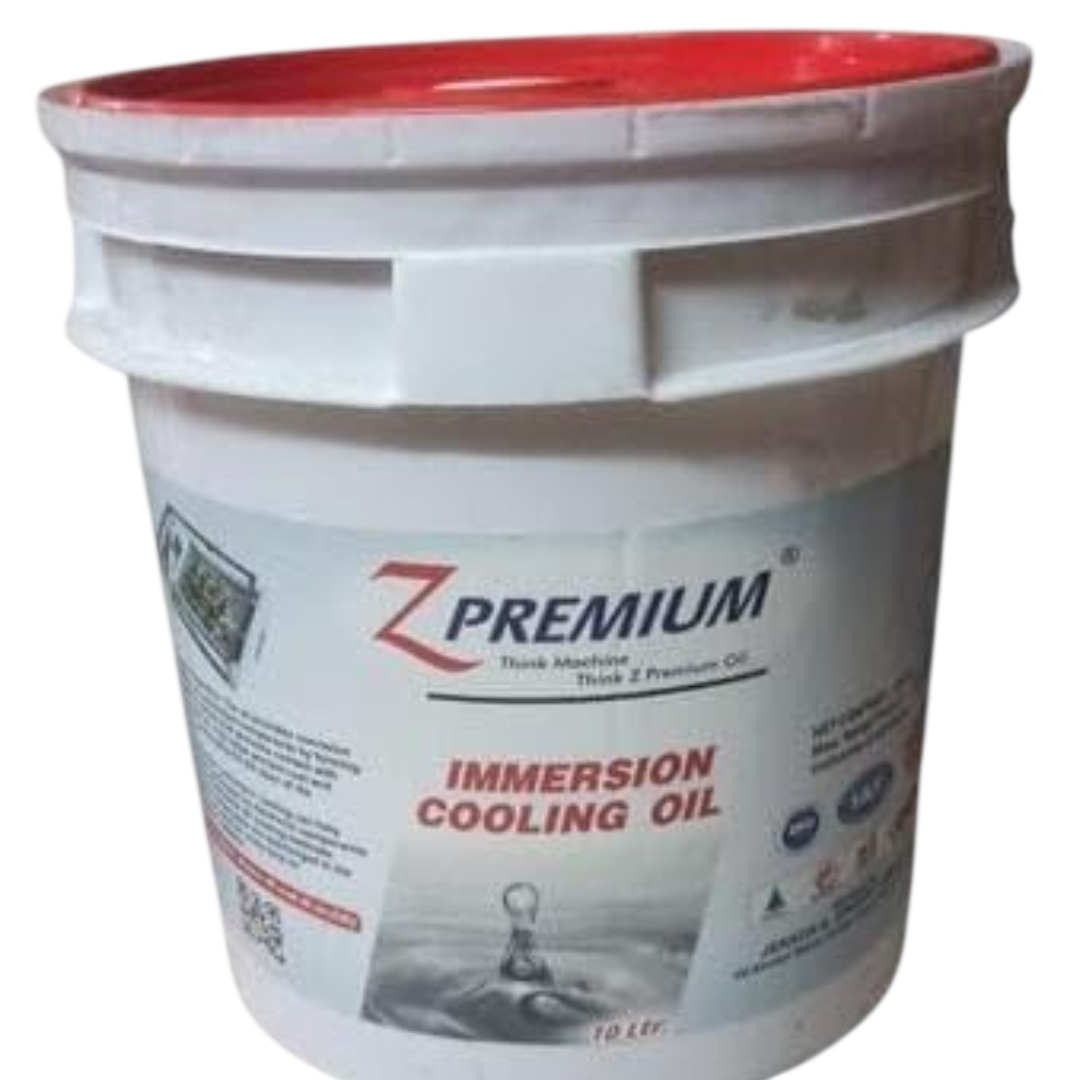}
        \caption{Mineral oil.}
    \end{subfigure}
    \hfill
    \begin{subfigure}[b]{0.17\linewidth}
        \centering
        \includegraphics[width=\linewidth]{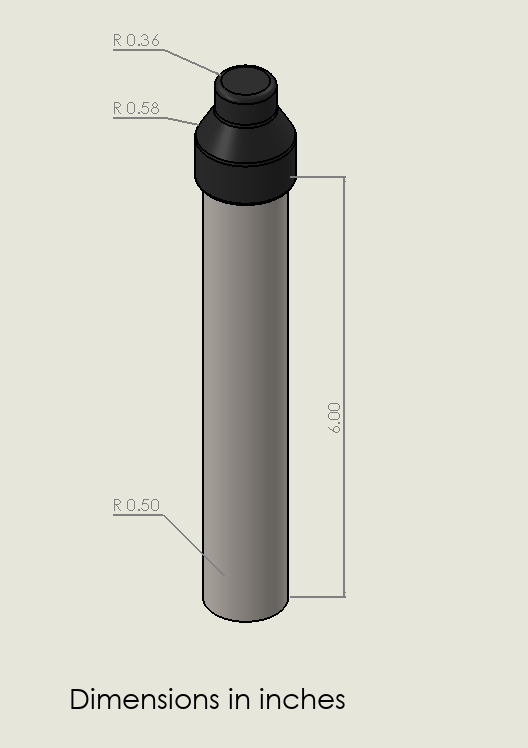}
        \caption{Heater (CAD).}
    \end{subfigure}
    \hfill
    \begin{subfigure}[b]{0.22\linewidth}
        \centering
        \includegraphics[width=\linewidth]{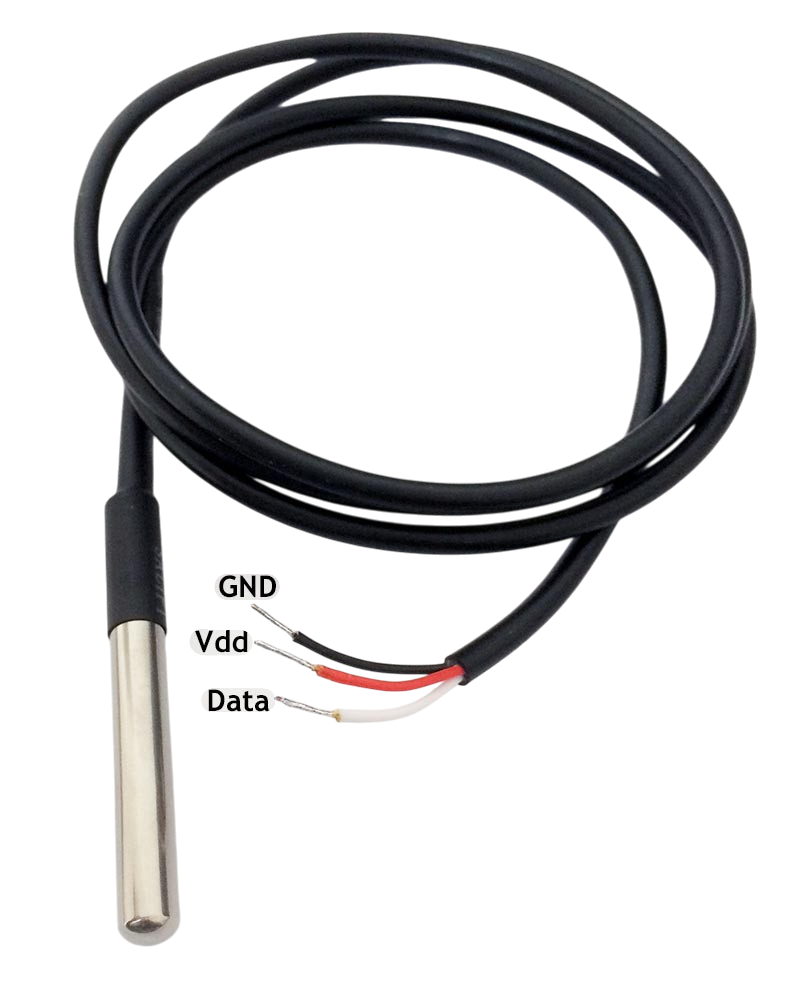}
        \caption{DS18B20 sensor.}
    \end{subfigure}
    \caption{Main hardware components: (a) glass immersion tank, (b) mineral oil, (c) titanium heating element and (d) DS18B20 temperature sensor.}
    \label{fig:hd_parts}
\end{figure}

The immersion tank serves as the primary enclosure of the system and is manufactured from glass. It is designed to safely contain the dielectric fluid and allow complete submersion of heat-generating components. This design consists of components that are affordable and locally accessible to achieve similar and comparable thermal readings at a fraction of the cost of commercial systems, which generally use expensive components. This prototype aims to replicate the core immersion cooling concept using locally accessible components compared to commercial systems. The immersion tank is shown in Fig.~\ref{fig:hd_parts}(a).

\subsection{Dielectric fluid}

A dielectric fluid is used as the heat transfer medium within the system. This fluid absorbs heat from the immersed components and transports it through the cooling loop. The mineral oil used in this system is shown in Fig.~\ref{fig:hd_parts}(b).

\subsection{Heat generation unit}

To replicate the thermal output of server hardware, an electrical heating element is partially immersed in the dielectric fluid. The heating element is shown in Fig.~\ref{fig:hd_parts}(c).

\subsection{Fluid circulation and heat rejection}

\begin{figure}[!htbp]
    \centering
    \begin{subfigure}[b]{0.30\linewidth}
        \centering
        \includegraphics[width=\linewidth]{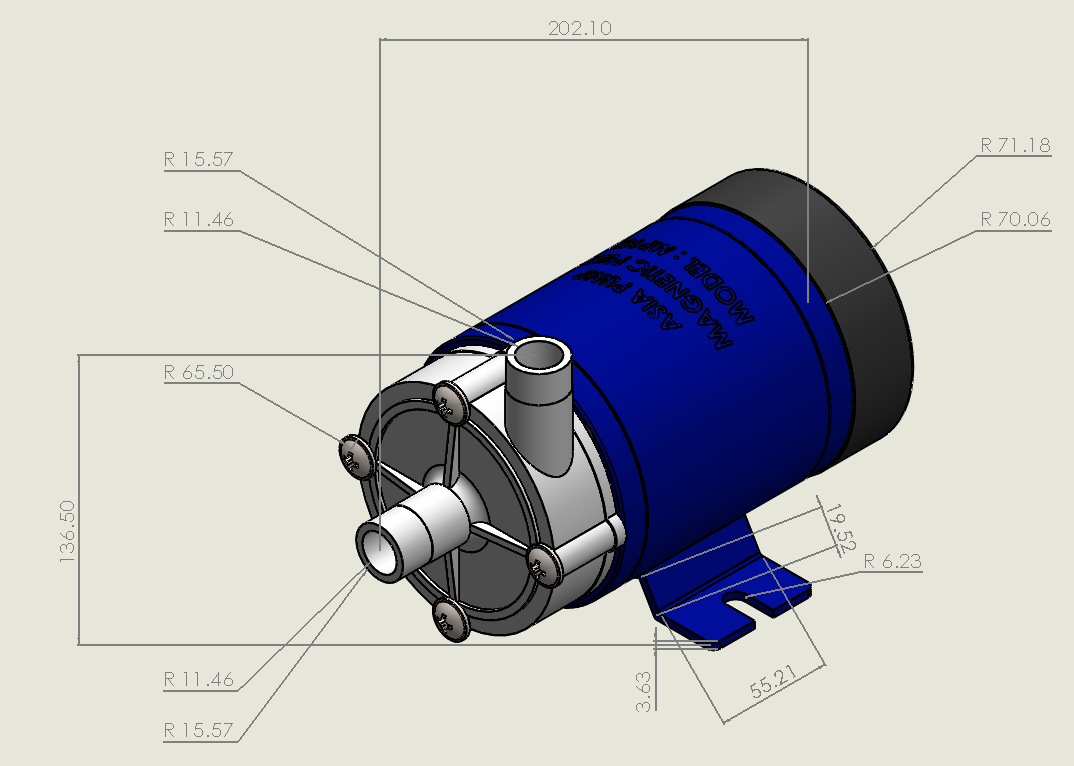}
        \caption{Fluid pump.}
    \end{subfigure}
    \hfill
    \begin{subfigure}[b]{0.30\linewidth}
        \centering
        \includegraphics[width=\linewidth]{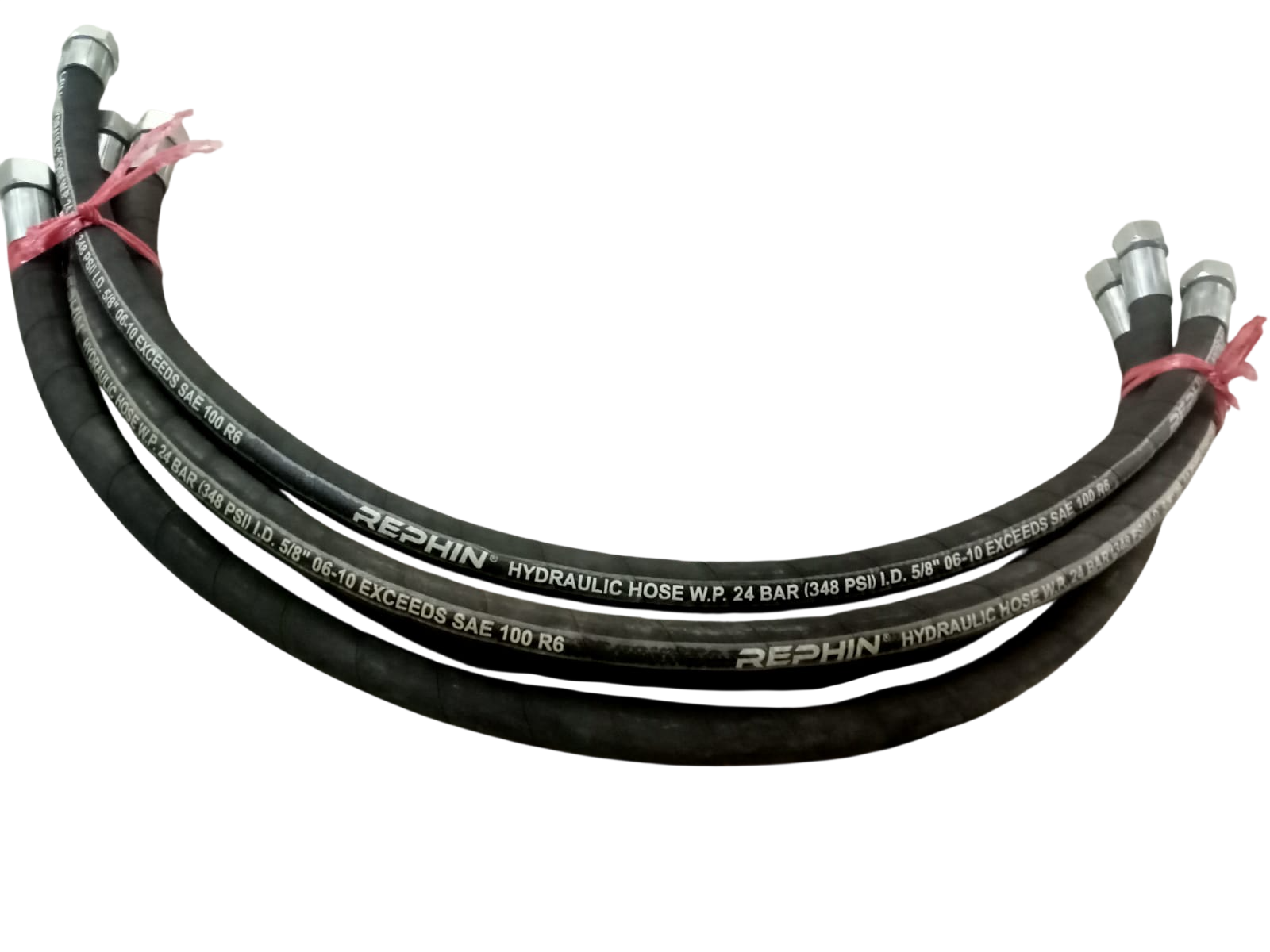}
        \caption{Hydraulic pipes.}
    \end{subfigure}
    \hfill
    \begin{subfigure}[b]{0.30\linewidth}
        \centering
        \includegraphics[width=\linewidth]{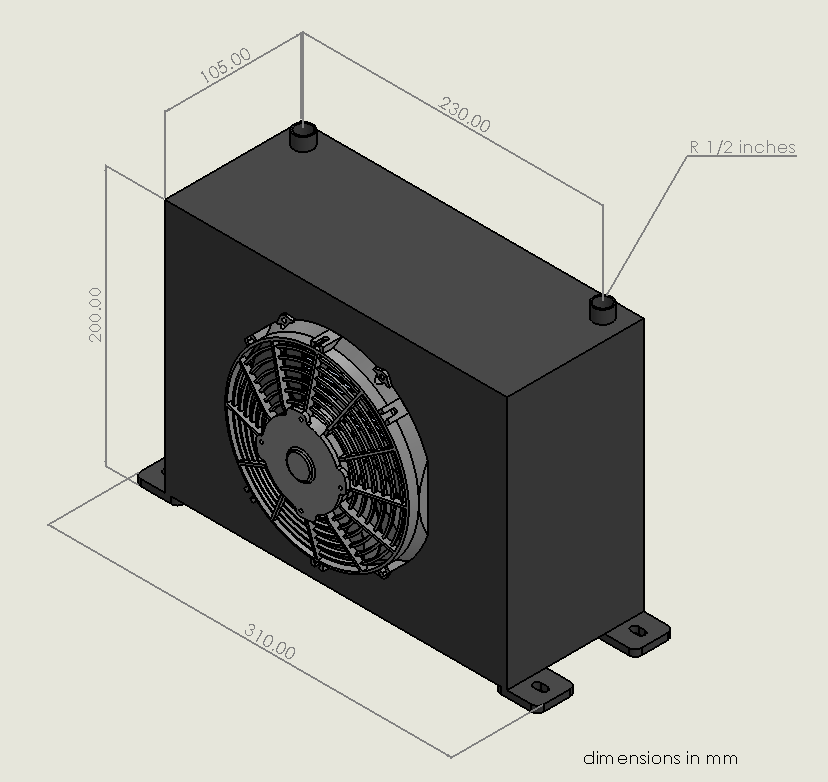}
        \caption{Oil cooler.}
    \end{subfigure}
    \caption{Fluid circulation and heat rejection components.}
    \label{fig:hd_circ}
\end{figure}

The circulation of the dielectric fluid is done using a pump, which maintains a continuous and uniform flow throughout the system. The heated fluid is directed through a radiator with fans acting as a heat exchanger, where thermal energy is dissipated to the ambient environment. Single-phase immersion cooling with oil coolant can provide effective heat transfer through natural and forced convection mechanisms \cite{ref3,ref7}. The pump, hydraulic pipes, and oil cooler are shown in Fig.~\ref{fig:hd_circ}(a), Fig.~\ref{fig:hd_circ}(b), and Fig.~\ref{fig:hd_circ}(c), respectively.

\subsection{Sensors and data acquisition}

Temperature monitoring is performed using temperature sensors placed within the immersion tank. These sensors continuously monitor the fluid temperature. The acquired data provides real-time insight into the thermal performance of the system and supports quantitative analysis of heat transfer efficiency. The DS18B20 temperature sensors used are shown in Fig.~\ref{fig:hd_parts}(d).

\subsection{Control and processing unit}

\begin{figure}[!htbp]
    \centering
    \includegraphics[width=0.6\linewidth]{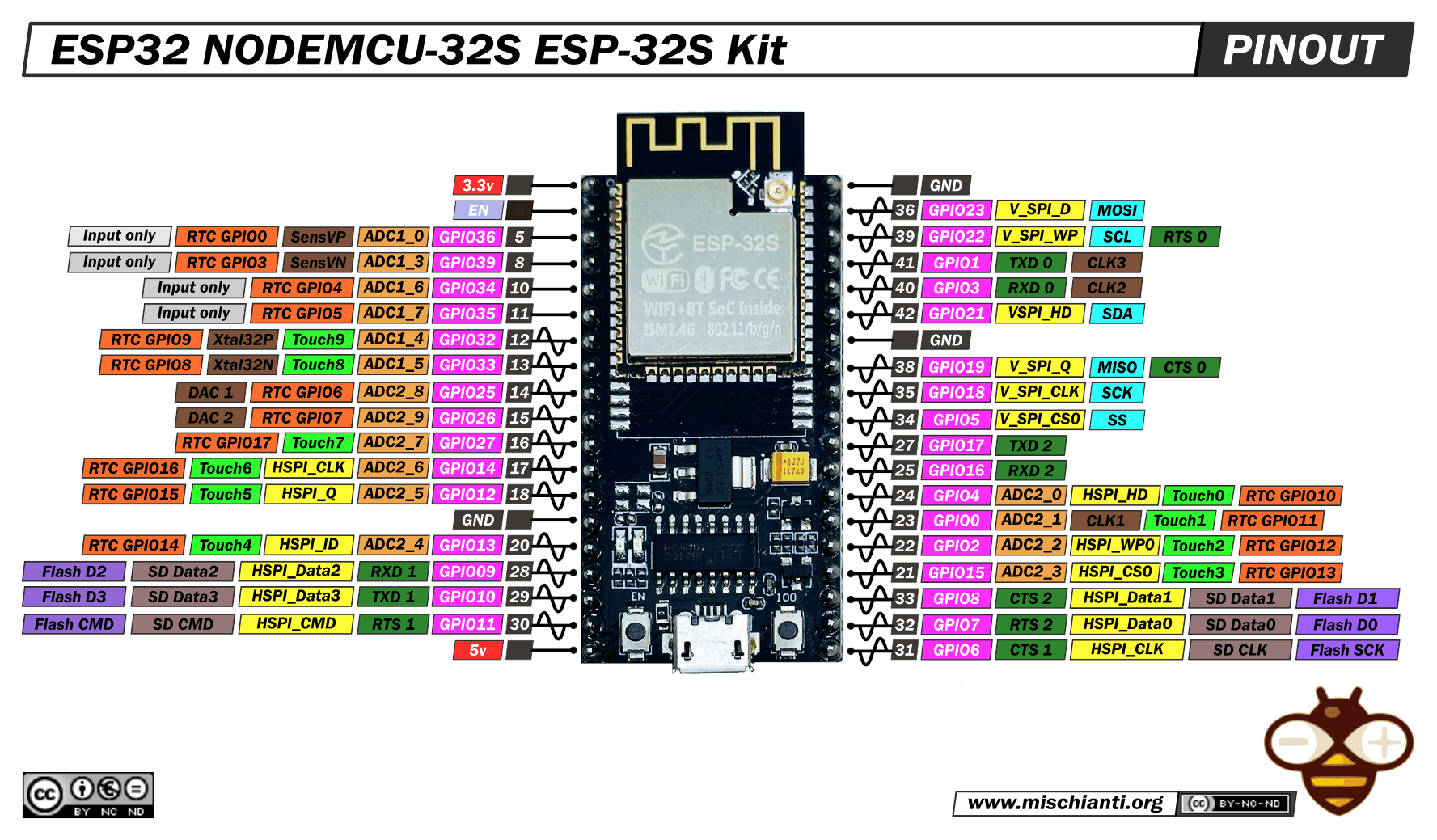}
    \caption{ESP32 NodeMCU-32S pinout. Image by R.~Mischianti (mischianti.org), licensed under CC BY-NC-ND and reproduced unmodified \cite{ref15}.}
    \label{fig:hd_esp32}
\end{figure}

An ESP32 microcontroller is used as the central control and processing unit. It continuously reads temperature values from the sensors placed in the immersion tank and processes this information in real time. Based on these readings, the ESP32 manages the operation of different components in the system. Since the AC components are to be managed safely, the system is incorporated with a small relay module that allows the ESP32 to switch all the AC components through control signals. This ensures safe operation when handling 230 V AC equipment while maintaining full electrical isolation in the dielectric environment. The pinout of the ESP32 development board used is shown in Fig.~\ref{fig:hd_esp32}.

\section{Design files}

\subsection{Design files description}

\begin{figure}[!htbp]
    \centering
    \includegraphics[width=0.75\linewidth]{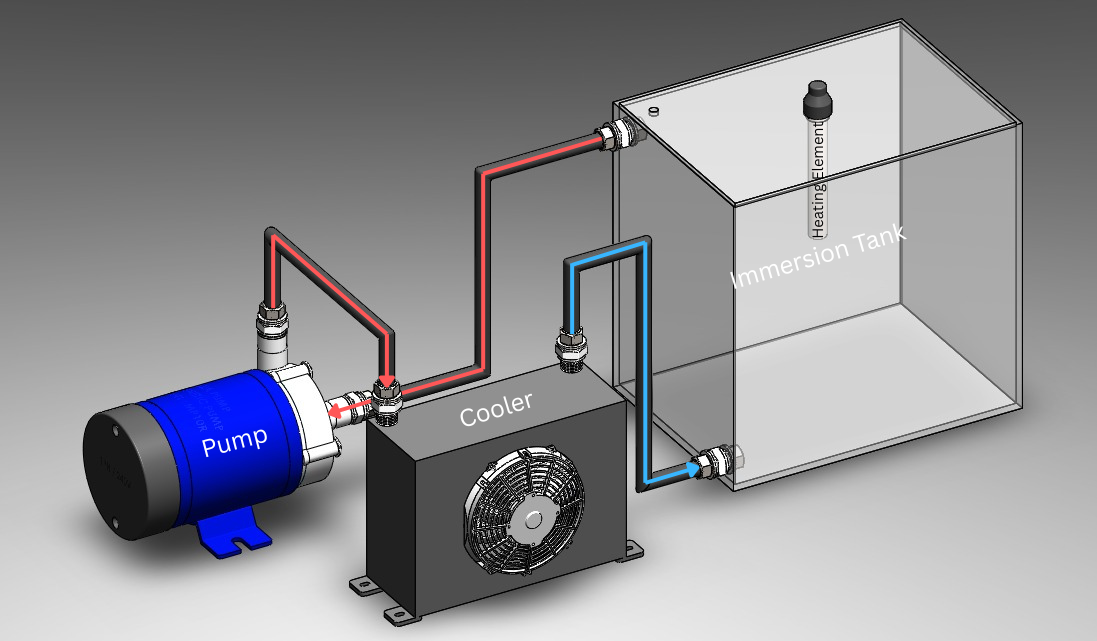}
    \caption{Final Assembly CAD.}
    \label{fig:final_assembly}
\end{figure}

The project package includes the CAD part and assembly files needed to understand and reproduce the system. Each file is listed in Table~\ref{tab:design_files} and described in Section~\ref{sec:design_summary}. The top-level assembly, which shows the complete layout of the tank, pump, radiator and interconnections, is shown in Fig.~\ref{fig:final_assembly}.

\section{Design files summary}
\label{sec:design_summary}

Table~\ref{tab:design_files} presents a comprehensive summary of all 
design files associated with this hardware, including CAD models and assembly files.

\begin{table}[!htbp]
\centering
\sloppy
\small
\caption{Design files summary. All files are released under the CC BY 4.0 International license and are hosted at \url{https://doi.org/10.17605/OSF.IO/SWZV7}.}
\label{tab:design_files}
\begin{tabular}{>{\raggedright\arraybackslash}p{0.30\linewidth}>{\raggedright\arraybackslash}p{0.13\linewidth}>{\raggedright\arraybackslash}p{0.30\linewidth}>{\raggedright\arraybackslash}p{0.13\linewidth}}
\toprule
\textbf{Design filename} & \textbf{File type} & \textbf{Design filename} & \textbf{File type} \\
\midrule
5 by 8 nut.SLDPRT & CAD Model & Mag drive pump.SLDPRT & CAD Model \\
ADAPTADOR MACHO X MACHO 5\_8 BSP.SLDPRT & CAD Model & New Tank.SLDPRT & CAD Model \\
Cooler Assembly.SLDASM & Assembly Model & Pipe1.SLDPRT & CAD Model \\
Cooler body.SLDPRT & CAD Model & Pipe2.SLDPRT & CAD Model \\
Dowty washer.SLDPRT & CAD Model & Pipe3.SLDPRT & CAD Model \\
Fan.SLDPRT & CAD Model & radiatorCoreDoublePass.SLDPRT & CAD Model \\
FINAL ASSEMBLY.SLDASM & Assembly Model & Tank + fittings + cover + heating element.SLDASM & Assembly Model \\
Heating element.SLDPRT & CAD Model & Tank cover.SLDPRT & CAD Model \\
Mag drive pump with fittings.SLDASM & Assembly Model &  &  \\
\bottomrule
\end{tabular}
\end{table}

All seventeen design files are released under the CC-By Attribution 4.0 
International license and are hosted at a single open-access repository at 
\url{https://doi.org/10.17605/OSF.IO/SWZV7}. 
The files span two SolidWorks formats --- individual part files (\texttt{.SLDPRT}) 
and multi-component assembly files (\texttt{.SLDASM}) --- and together cover every 
subsystem of the hardware, from small individual fasteners to the complete 
integrated system.

The mechanical and sealing hardware is captured in three part files. The 
\texttt{5 by 8 nut.SLDPRT} and the \texttt{ADAPTADOR MACHO X MACHO 5\_8 BSP.SLDPRT} 
model the threaded fasteners and male-to-male BSP adapter used at all hydraulic 
junctions throughout the system. The \texttt{Dowty washer.SLDPRT} models the 
face-sealing washers that serve as the primary leak-prevention element at every 
pipe and port connection.

The heat rejection subsystem is documented across four files. The radiator 
body is modeled standalone in \texttt{Cooler body.SLDPRT}, the cooling fan in 
\texttt{Fan.SLDPRT}, and the double-pass radiator core geometry --- responsible for 
dissipating heat to the ambient environment --- in 
\texttt{radiatorCoreDoublePass.SLDPRT}. All three are brought together in the 
\texttt{Cooler Assembly.SLDASM}, which represents the complete heat-rejection unit 
as it is physically assembled.

The fluid circulation hardware consists of two pump files and three pipe 
files. The magnetic drive pump is first modeled as a standalone part in 
\texttt{Mag drive pump.SLDPRT}, and then integrated with its inlet and outlet 
fittings in the \texttt{Mag drive pump with fittings.SLDASM} assembly. The three 
hydraulic hoses that route the dielectric fluid between the tank, pump, and 
radiator are individually captured in \texttt{Pipe1.SLDPRT}, \texttt{Pipe2.SLDPRT}, 
and \texttt{Pipe3.SLDPRT}.

The immersion tank and heating subsystem is represented by four files. The 
tank body is modeled in \texttt{New Tank.SLDPRT} and its removable lid in 
\texttt{Tank cover.SLDPRT}. The 500~W titanium heating element that simulates 
server thermal output is captured in 
\texttt{Heating element.SLDPRT}. These, along with the bulkhead fittings, are 
combined into the \texttt{Tank + fittings + cover + heating element.SLDASM} 
assembly file, which represents the complete tank subassembly prior to filling.

Finally, \texttt{FINAL ASSEMBLY.SLDASM} integrates every subsystem into the 
complete system model, directly corresponding to the physical layout shown in 
Fig.~\ref{fig:final_assembly}, and serves as the top-level reference for 
understanding how all components interconnect.

\section{Bill of materials}

Table~\ref{tab:bom} presents the complete bill of materials for the 
immersion cooling system, listing all components with their costs, sources, and 
material types.

\begin{table}[!htbp]
\centering
\sloppy
\small
\caption{Bill of Materials}
\label{tab:bom}
\begin{tabular}{>{\raggedright\arraybackslash}p{0.155\linewidth}>{\raggedright\arraybackslash}p{0.11\linewidth}>{\raggedright\arraybackslash}p{0.085\linewidth}>{\raggedright\arraybackslash}p{0.085\linewidth}>{\raggedright\arraybackslash}p{0.085\linewidth}>{\raggedright\arraybackslash}p{0.15\linewidth}>{\raggedright\arraybackslash}p{0.125\linewidth}}
\toprule
\textbf{Designator} & \textbf{Component} & \textbf{Number} 
& \textbf{Cost per unit (INR)} & \textbf{Total cost (INR)} 
& \textbf{Source of materials} & \textbf{Material type} \\
\midrule
New Tank, Tank cover & Tank & 1 & 2100 & 2100 
& Fish Kingdom Officials, Mandi, HP & Inorganic \\
Mag drive pump & Motor Pump & 1 & 4760 & 4760 
& Taha Industries & Metal \\
Cooler Assembly & Oil Cooler & 1 & 9205 & 9205 
& Ace Automation Engineers & Metal \\
Heating element & Heating element & 1 & 944 & 944 
& Shree Bhavna Sales & Metal \\
- & Mineral Oil & 35 L & 416 & 14,570 
& Jenson \& Jenson Lubricant & Organic \\
Pipe1, Pipe2, Pipe3 & Pipes & 3 & 335 & 1005 
& Indian Rubber and Hydraulics Co. & Composite \\
- & Sensors & 3 & 266 & 799 
& Moxie Supply Pvt. Ltd., Macfos Ltd. & Semiconductor \\
5 by 8 nut, ADAPTADOR MACHO X MACHO 5 8 BSP & Connectors & 6 & 102 & 614 
& Indian Rubber and Hydraulics Co. & Metal \\
- & MCU and Wiring & - & - & 911 
& A2Z, IIT Mandi & Semiconductor \\
- & Sealant and Adhesives & 2 & 205 & 410 
& Raja Traders, Delhi & Polymer \\
\midrule
\multicolumn{4}{l}{\textbf{Grand Total}} & \textbf{35318} & & \\
\bottomrule
\end{tabular}
\end{table}

The system was assembled for a grand total of INR 35,318, sourced 
entirely from vendors accessible within India, underscoring the 
project's commitment to local reproducibility. The ten line items span inorganic, 
metallic, composite, semiconductor, and polymer material categories, and were 
procured from seven different suppliers across Himachal Pradesh, Delhi, and IIT 
Mandi's own campus store.

The single largest cost is the mineral oil, at INR 14,570 for 35~litres 
(INR 416 per litre from Jenson \& Jenson Lubricant), which alone accounts for 
roughly 41\% of the total budget. This reflects the 
substantial fluid volume required to immerse the heating element to the required depth and fully submerge the sensor
probes within the 350~$\times$~250~$\times$~350~mm tank. The next two major items 
are the oil cooler (Cooler Assembly, INR 9,205 from Ace Automation 
Engineers) and the magnetic drive pump (INR 4,760 from Taha Industries), 
which together form the active heat-rejection loop and account for a further 40\% 
of the total. The glass tank and removable cover, procured from Fish 
Kingdom Officials in Mandi, HP, cost INR 2,100 for the complete enclosure.

The remaining costs are distributed across supporting subsystems. The three 
hydraulic pipes (Pipe1--Pipe3) cost INR 1,005 in total (INR 335 each) 
from Indian Rubber and Hydraulics Co., the same vendor from which the six BSP connectors --- the 5/8 hex nuts and the ADAPTADOR MACHO X MACHO fittings 
--- were procured for INR 614 (INR 102 per connector). The heating 
element, a 500~W titanium rod from Shree Bhavna Sales used to simulate server thermal output, cost INR 944. The three DS18B20 digital temperature sensors, sourced jointly from Moxie Supply Pvt.\ Ltd.\ and 
Macfos Ltd., added INR 799 to the total at INR 266 per sensor.

Finally, the microcontroller, relay module, and wiring (MCU and Wiring) 
were assembled from components available at A2Z, IIT Mandi's campus store for 
INR 911, and two units of Anabond 666 RTV silicone sealant and adhesive
from Raja Traders, Delhi, rounded out the bill at INR 410 (INR 205 per unit). Thermal Grizzly Hydronaut 
\cite{ref16} was selected as the thermal interface material at the heater 
sensor contact due to its silicone-free formulation --- precluding any risk of 
silicone leaching into the mineral oil --- and its rated operating range of 
$-200$~\textdegree{}C to $+350$~\textdegree{}C \cite{ref16}, which encompasses the observed heater 
surface temperatures. The 
entire system therefore falls well within its target budget, and every component 
can be individually replaced or upgraded without specialized procurement channels.

\section{Build instructions}

\subsection{Tank fabrication and fittings}

The immersion cooling tank was customized to accommodate a standard mATX motherboard (305 x 244 mm) to enable future retrofitting of live compute hardware. The tank is constructed with 5 mm thick glass panels, with internal dimensions of 350 mm (L) x 250 mm (W) x 350 mm (H). A 500 W titanium heating element (6-inch length, 1-inch diameter) is used to replicate the server's heat. A small aperture is made on top of the removable top glass panel, which allows routing of the sensor wires. A circular opening is cut at the center of the removable glass panel to insert the heating element vertically. The 350 x 250 mm side face includes two holes diagonally for the hot fluid outlet and cool fluid inlet.

Piping and sealing specifications:

\begin{enumerate}
\item Hoses: 1m Rephin Hydraulic hose SAE100 R6 5/8" ID with 5/8" BSPP female swivel fittings crimped on ends.
\item Cooler to pipe junction: 5/8" to 1/2" BSPP reducing adapter + 1/2" Dowty washer.
\item Pump and tank ports: 5/8" BSPP hex nipples + 5/8" Dowty washers.
\item Mechanical reinforcement: 5/8" hex nuts on the inner side of the tank wall.
\item Sealing: All fittings use Dowty washers as the primary seal. Industrial-grade RTV silicone sealant Anabond 666 is applied externally for secondary sealing.
\end{enumerate}

\begin{figure}[!htbp]
\centering
\begin{subfigure}{0.45\textwidth}
    \includegraphics[width=\linewidth]{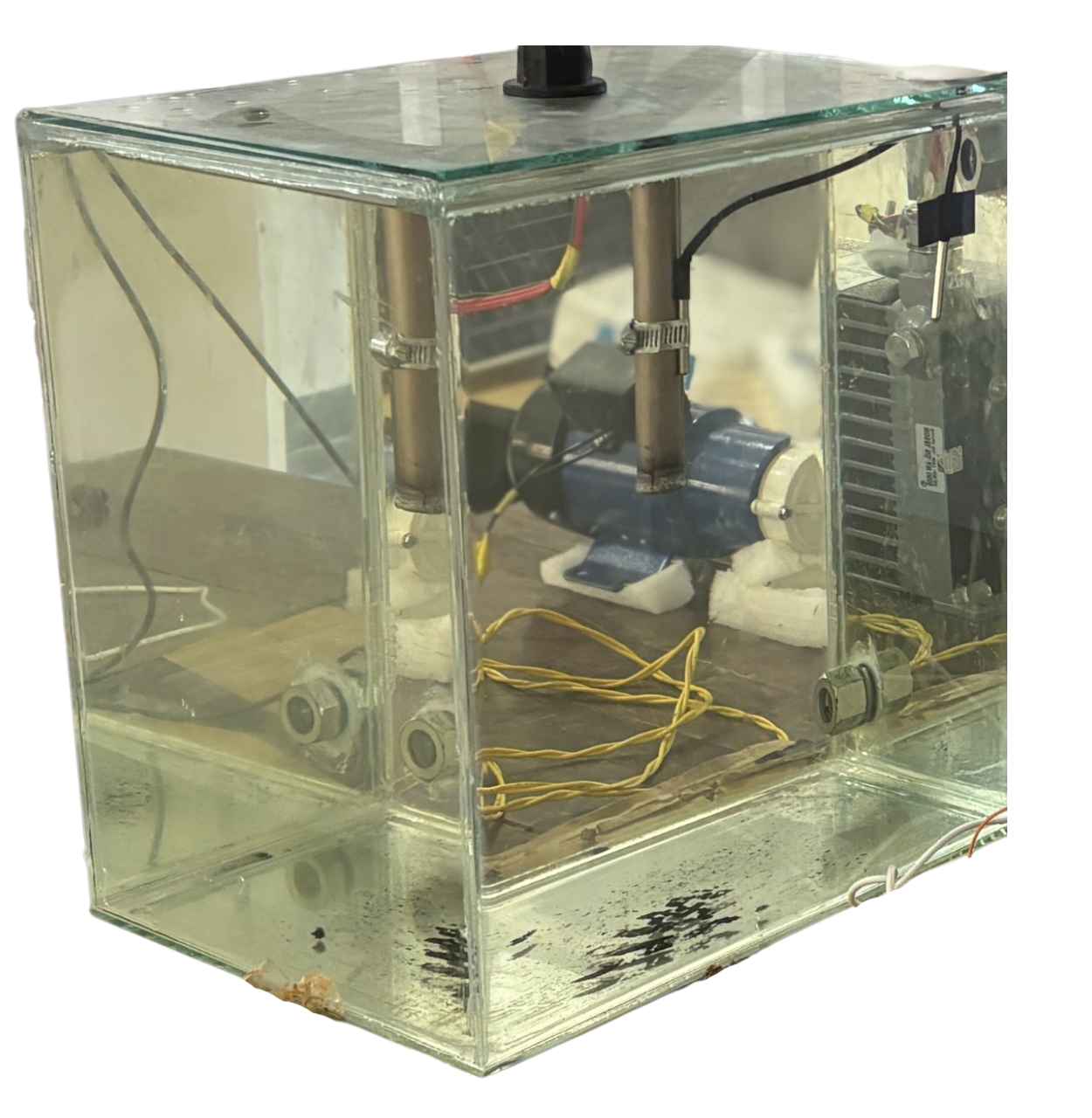}
    \caption{Immersion Tank.}
    \label{fig:tank}
\end{subfigure}
\hfill
\begin{subfigure}{0.3\textwidth}
    \includegraphics[width=\linewidth]{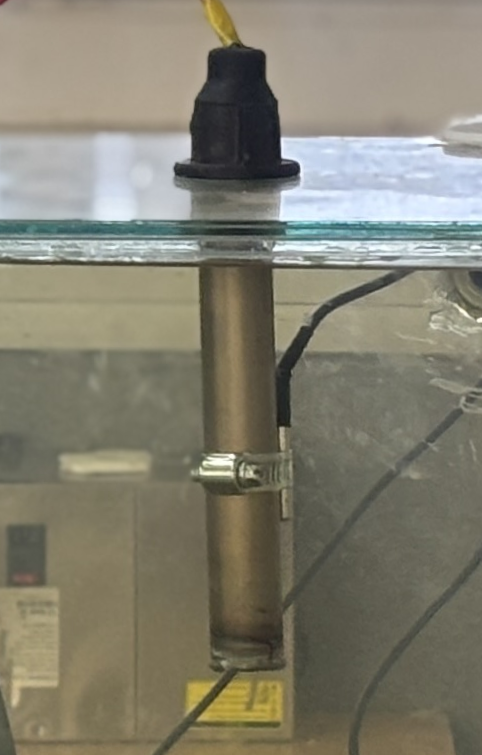}
    \caption{Heating Element.}
    \label{fig:heater}
\end{subfigure}
\caption{Tank assembly and heating element.}
\label{fig:tank_assembly}
\end{figure}

\indent \indent The image of our customized tank is shown in Fig.~\ref{fig:tank_assembly}(a) and the heating element in Fig.~\ref{fig:tank_assembly}(b).

\subsection{Cooling loop assembly}

Components:
\begin{enumerate}
\item Pump: A PMD-15 Magnetic Drive Pump (15~LPM). This was selected for its 
compatibility with mineral oil. It is shown in 
Fig.~\ref{fig:cooling_components}(b).
\item Radiator: An Air-Cooled Oil Cooler. As shown in 
Fig.~\ref{fig:cooling_components}(a), this was chosen for its 
compatibility with mineral oil and its flow rate being compatible with the pump.
\end{enumerate}

Assembly:
\begin{enumerate}
\item The port at the top of the tank is connected to the pump's inlet via a hose.
\item The pump's outlet is connected to the radiator's inlet.
\item The radiator's outlet is connected to the port at the bottom of the tank 
via a hose.
\item All joints are fastened using the specified BSP fittings and Dowty washers 
to ensure reliable sealing.
\end{enumerate}

\begin{figure}[!htbp]
\centering
\begin{subfigure}{0.3\textwidth}
    \includegraphics[width=\linewidth]{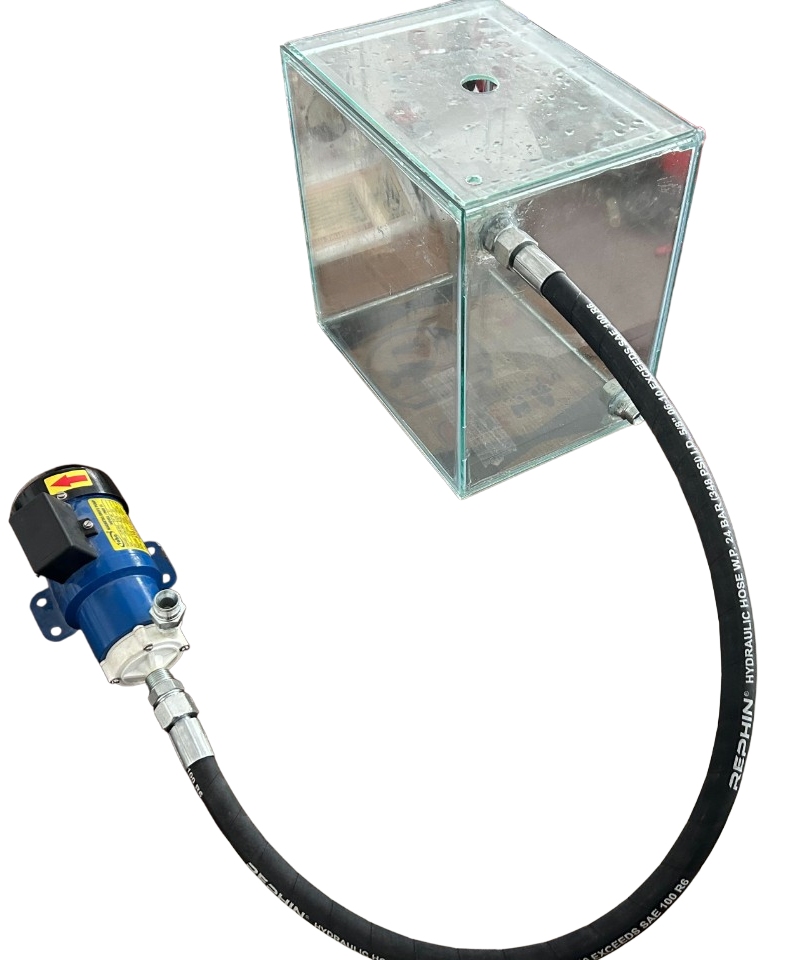}
    \caption{Step 1: Top tank port connected to pump inlet via hydraulic hose.}
    \label{fig:assembly_step1}
\end{subfigure}
\hfill
\begin{subfigure}{0.3\textwidth}
    \includegraphics[width=\linewidth]{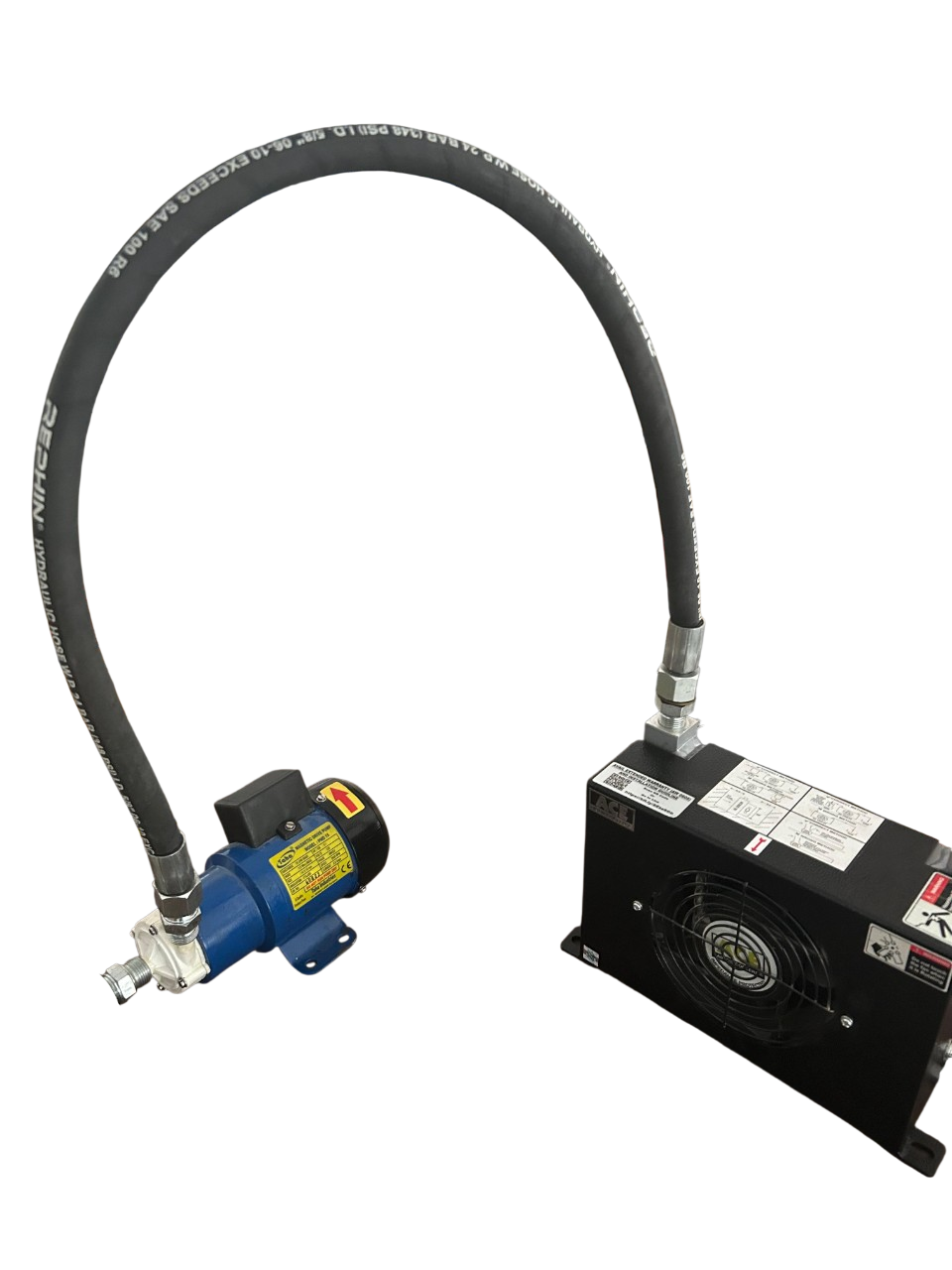}
    \caption{Step 2: Pump outlet connected to radiator inlet.}
    \label{fig:assembly_step2}
\end{subfigure}
\hfill
\begin{subfigure}{0.3\textwidth}
    \includegraphics[width=\linewidth]{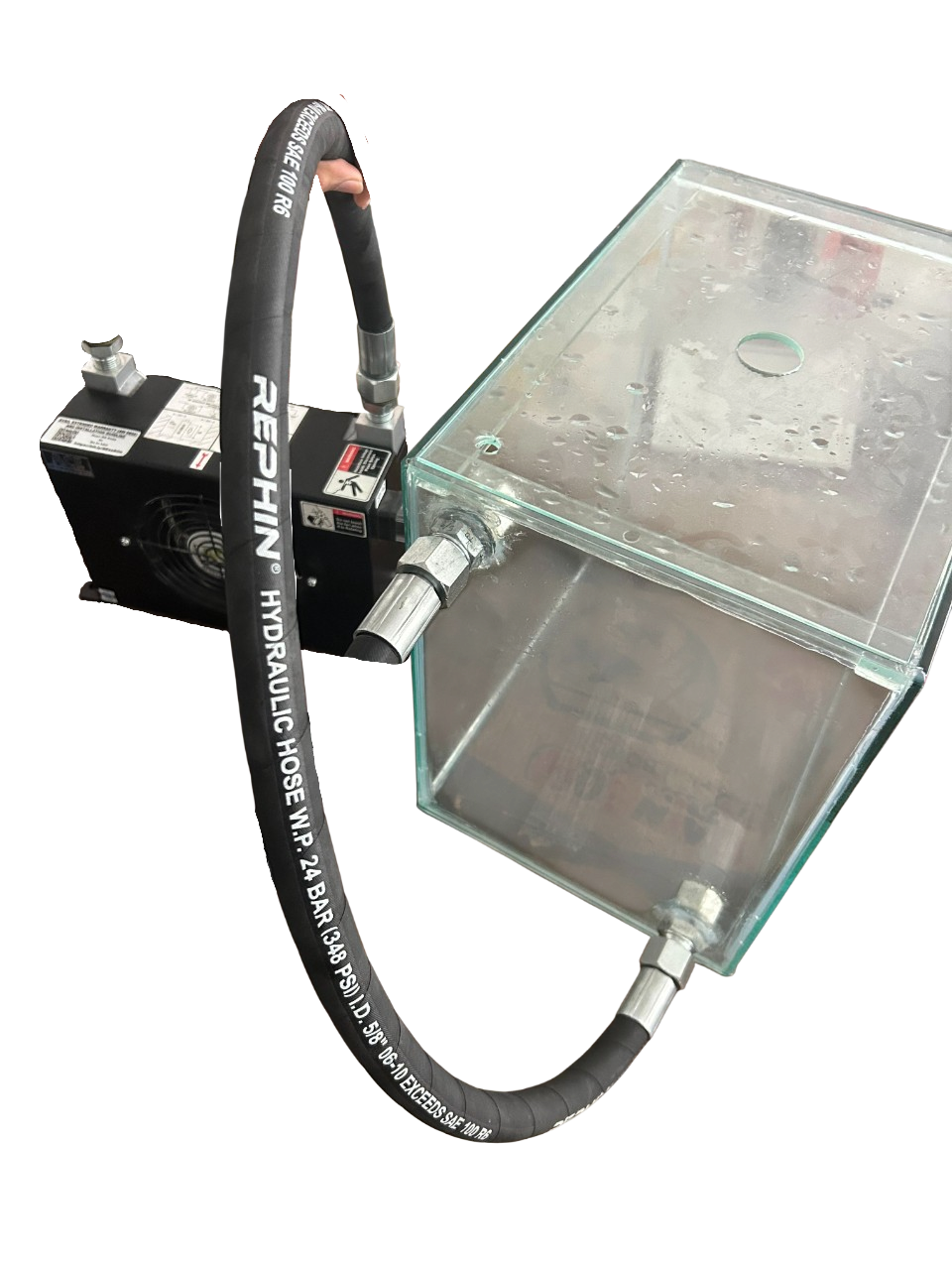}
    \caption{Step 3: Radiator outlet connected to bottom tank port via hydraulic 
    hose.}
    \label{fig:assembly_step3}
\end{subfigure}
\caption{Cooling loop assembly sequence.}
\label{fig:assembly_steps}
\end{figure}

The three assembly steps above are illustrated sequentially in 
Fig.~\ref{fig:assembly_steps}, showing the hydraulic connections between 
the tank, pump, and radiator as they are established.

The radiator and pump used in this project are shown in 
Fig.~\ref{fig:cooling_components}:

\begin{figure}[!htbp]
\centering
\begin{subfigure}{0.45\textwidth}
    \includegraphics[width=\linewidth]{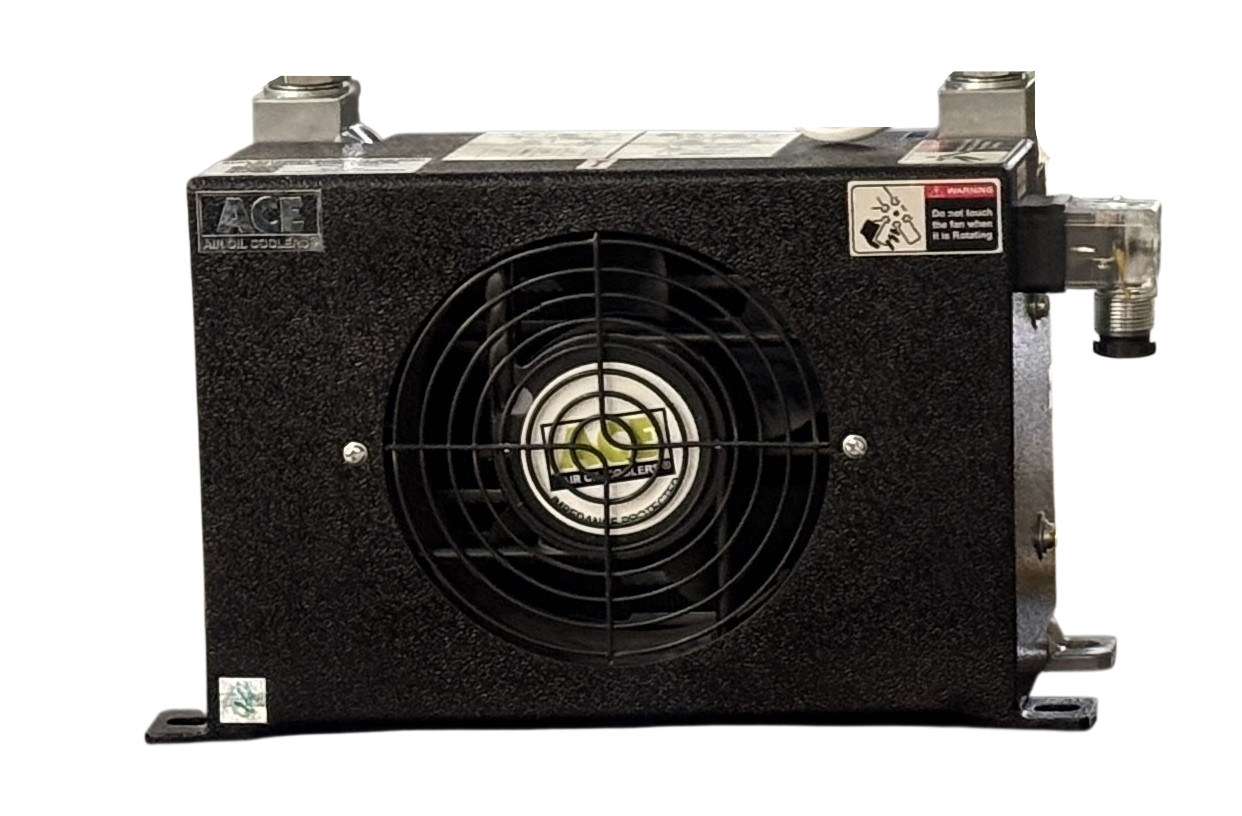}
    \caption{Cooler.}
    \label{fig:cooler}
\end{subfigure}
\hfill
\begin{subfigure}{0.45\textwidth}
    \includegraphics[width=\linewidth]{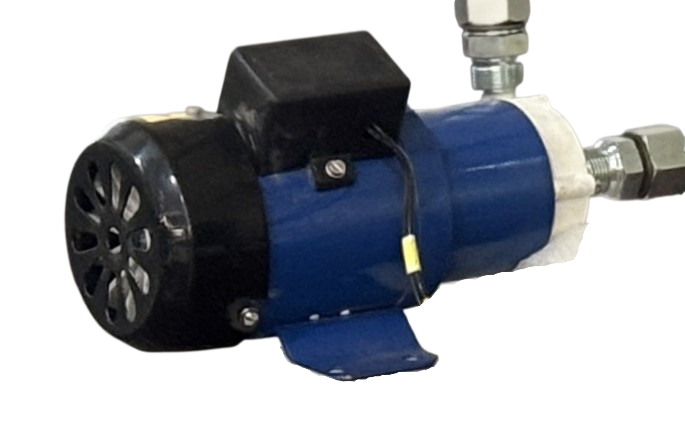}
    \caption{Pump.}
    \label{fig:pump}
\end{subfigure}
\caption{Cooling Loop Components.}
\label{fig:cooling_components}
\end{figure}

\subsection{Electronics and sensor integration}

The monitoring of temperatures from multiple points in the setup is done using an ESP32 microcontroller. Three DS18B20 digital temperature sensors are used to measure the temperatures.

AC Power and Relay Control: An HW-314 4-channel relay module is used to safely switch the 230 V AC components. The ESP32 controls the relays using 3.3 V logic, while the relay coils are powered by a separate 5 V source.

\begin{enumerate}
\item GPIO16 $\rightarrow$ IN1 (Pump Control).
\item GPIO17 $\rightarrow$ IN2 (Cooler Fan Control).
\item GPIO18 $\rightarrow$ IN3 (Heater Control).
\end{enumerate}

Sensor Integration:

\begin{enumerate}
\item Heater sensor: It is directly clamped to the heating element to measure 
its surface temperature (T\textsubscript{heater}). A thin layer of thermal 
paste \cite{ref16} is applied at the contact interface prior to clamping to 
minimise contact resistance. It is connected to the GPIO25.
\item Fluid sensor: The sensor is located inside the tank, affixed near the outlet of the warm fluid on the wall of the tank to monitor the bulk fluid temperature (T\textsubscript{fluid}) and is connected to GPIO26.
\item Ambient sensor: The sensor is located outside of the setup to measure the room temperature (T\textsubscript{ambient}) and is connected to GPIO27.
\end{enumerate}

\begin{figure}[!htbp]
    \centering
    \includegraphics[width=0.9\linewidth]{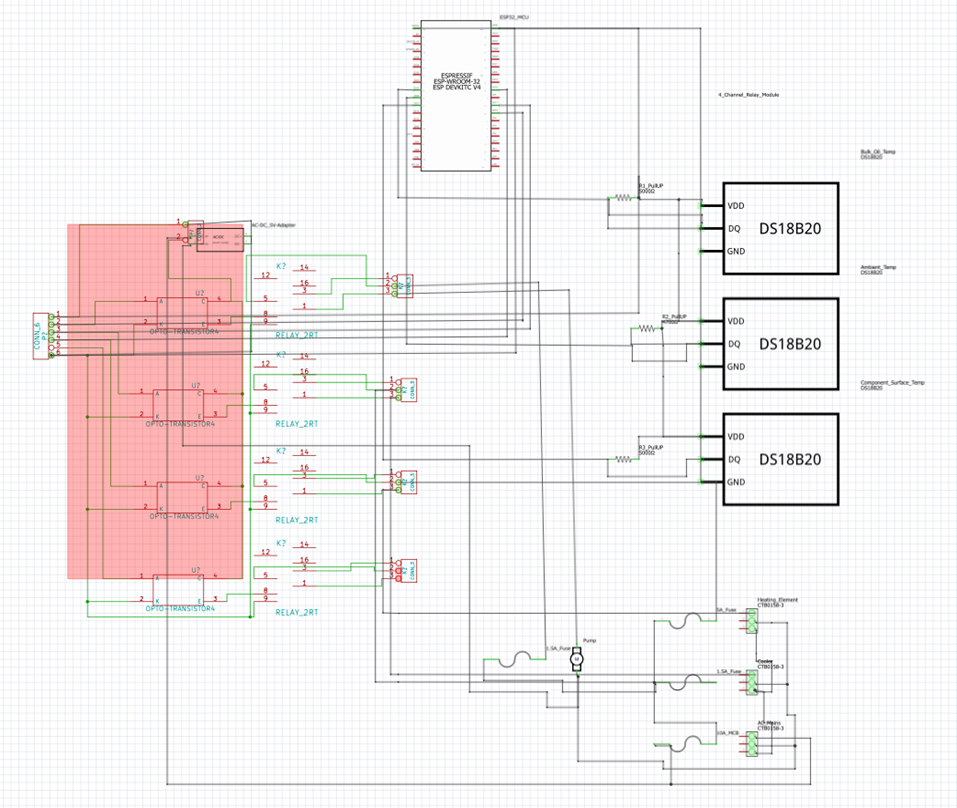}
    \caption{Schematic Diagram of Circuit.}
    \label{fig:circuit}
\end{figure}

\indent \indent Refer to Fig.~\ref{fig:circuit} for the schematic diagram of the circuit. Each of the DS18B20 sensors requires a minimum of a 4.7 k$\Omega$ pull-up resistor between the DQ line and the 3.3 V rail of ESP32 to ensure stable communication. The firmware is then used for logging and monitoring; it is provided in \hyperref[sec:appendixA]{Appendix A}.

\subsection{Final assembly and coolant filling}

After fixing all the loops and sensors, the tank is then filled with Z Premium Immersion Cooling Oil. This is a mineral-based dielectric oil with a high flash point of 200\textdegree{}C and a density of 808 kg/m\textsuperscript{3} \cite{ref13}. The fluid level must submerge the heating element to a certain level and the sensor probes completely. The complete assembled system is shown in Fig.~\ref{fig:setup}(a).

\subsection{AC Power Distribution and Relay Wiring}

The system uses a single-phase 230~V AC supply to power the heater, pump, and radiator fan. The live line passes through a 10~A MCB and is then divided into three branches, each protected by an individual fuse: 5~A for the heater and 1.5~A slow-blow fuses for the pump and fan. The fuse outputs are connected to the common (COM) terminals of the relay channels, and the normally open (NO) terminals are connected to the live terminals of the respective loads.

\begin{figure}[!htbp]
    \centering
    \includegraphics[width=0.6\linewidth]{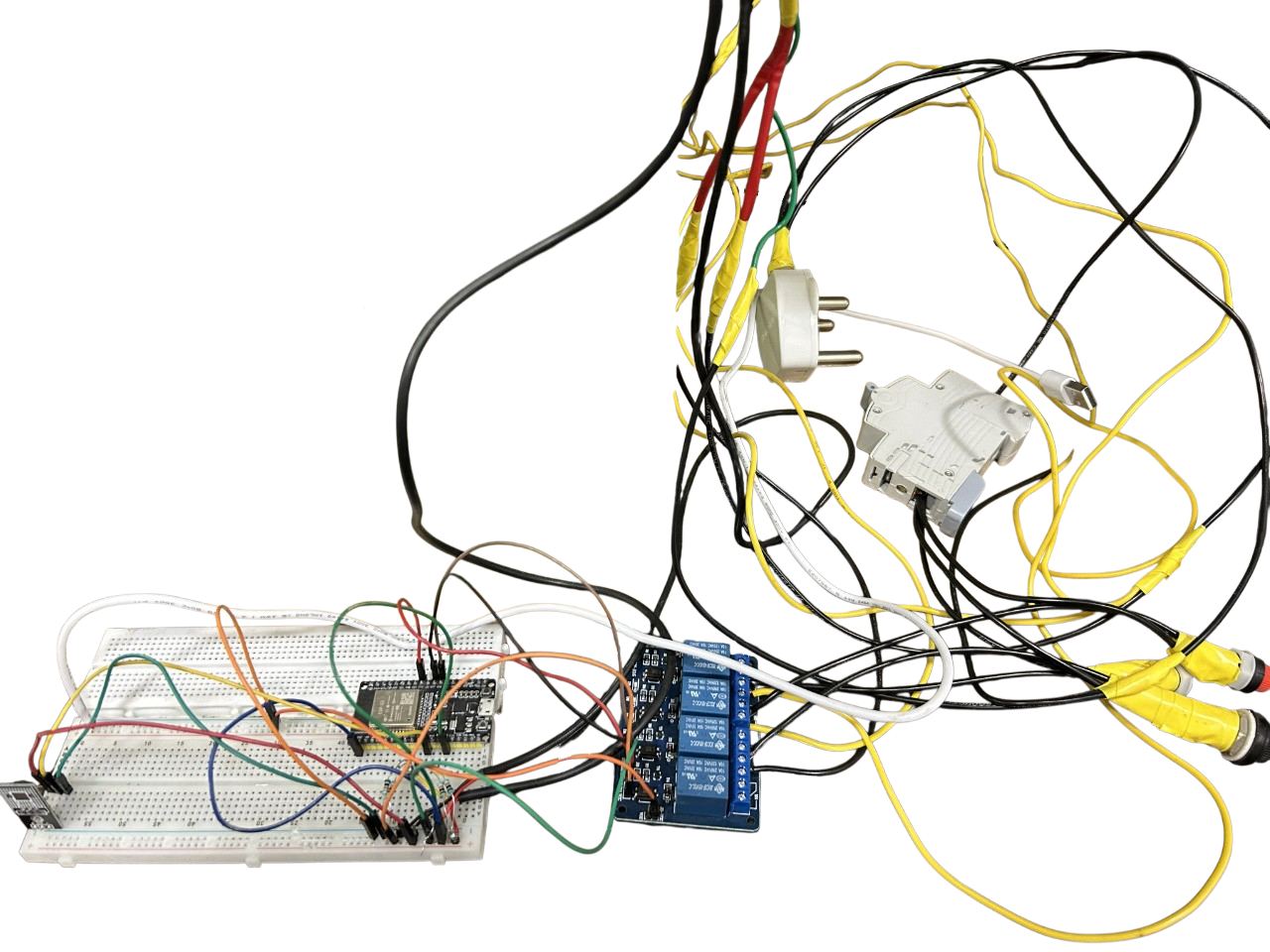}
    \caption{AC power distribution and relay wiring.}
    \label{fig:wiring}
\end{figure}

The neutral line is connected to a common neutral bus, while the protective earth is connected to a common earth bus and bonded to all metallic parts. The ESP32 controls the relay module using 3.3~V logic, with the relay coils powered separately at 5~V. This arrangement provides isolated switching of the AC loads while maintaining overcurrent protection and proper grounding. The complete wiring arrangement is shown in Fig.~\ref{fig:wiring}.

\subsection{Control strategy and algorithm}

The system operates under closed-loop temperature-based control on the ESP32 microcontroller. Three DS18B20 sensors monitor the heater surface, bulk fluid, and ambient temperatures, with heater temperature as the primary control variable.

The control algorithm compares heater temperature against predefined thresholds. Exceeding the lower threshold activates the circulation pump; reaching the upper threshold engages the radiator fan. Both deactivate automatically once temperatures fall below their respective lower thresholds. Manual override is available via serial input, and a safety cutoff disables the heater if a critical temperature limit is exceeded.

\begin{figure}[!htbp]
    \centering
    \includegraphics[width=0.42\linewidth]{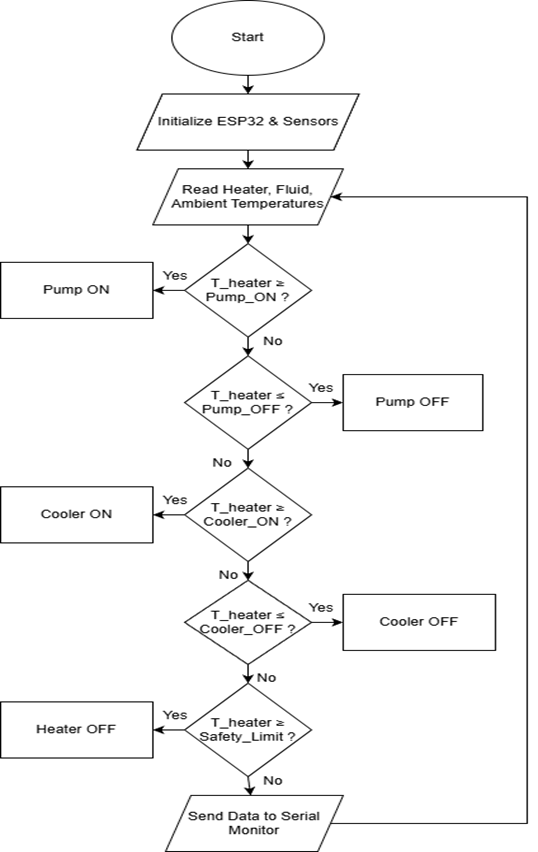}
    \caption{Operation flowchart of the system.}
    \label{fig:flowchart}
\end{figure}

The complete firmware implementing this logic is provided in \hyperref[sec:appendixA]{Appendix A}. The operational sequence of initialization, steady-state operation, and safe shutdown is summarized in the flowchart presented in Fig.~\ref{fig:flowchart}.

\section{Operation instructions}

\subsection{Safety precautions}

\begin{enumerate}
\item Make sure that all of the AC wiring is insulated and grounded.
\item Do not touch the heating element or the tank fittings while they are running or right after they are done, as they can get hotter than 100 \textdegree{}C and stay that way for several minutes.
\item Before you turn on the pump, make sure it is already primed. Also, make sure the coolant is flowing correctly to keep it from overheating.
\item Check the tank, hoses, and bulkhead fittings for oil leaks before each run. Even small leaks can get worse when they are hot and under pressure.
\item Place the setup on a flat, strong surface, and make sure the tank isn't too full so that the hot oil has room to expand.
\item If the pump flow slows down or stops, stop the experiment right away. The heater can get too hot without circulation.
\item Always have a fire extinguisher close by, and never leave the system running without supervision, especially when there is AC power and high temperatures.
\end{enumerate}

\subsection{Pre-operation checks}

\begin{enumerate}
\item Make sure all connections are tight with no leaking oil.
\item Check the fluid level so the heating element is submerged at the required level and the sensors are fully submerged.
\item Connect the ESP32 microcontroller via USB to the computer.
\end{enumerate}

\subsection{Startup procedure}

\begin{enumerate}
\item Monitor initialization: Open the Arduino IDE on the computer, select the COM port for the ESP32 and open the Serial Monitor. Set the baud rate to 115200. The ESP32 will stream live temperature data from the sensors.
\item Power application: Connect the system mains (plug) and switch the main MCB / fuses on. On power-up, nothing is energised by default. The pump and cooler run automatically according to the heater surface temperature thresholds; they will turn on as the heater temperature rises.
\item Heater activation: The heating element is manual --- enable it via the serial command provided as comments in the code in \hyperref[sec:appendixA]{Appendix A}. After the heater is enabled, monitor the serial stream until steady state is reached (steady-state flag in the output). The system is now active under automated pump/cooler control.
\end{enumerate}

\subsection{Shutdown procedure}

\begin{enumerate}
\item Heater deactivation: Turn the heater off first (serial command). The pump and cooler will remain under automatic control and will continue running as needed to cool the system.
\item System cool-down: Allow the pump and radiator to run for $\sim$5--10 minutes (or until temperature readings reach safe levels).
\item System power-down: Once temperatures are within the safe range, you may stop the pump and radiator (either wait for automatic turn-off per thresholds or use manual commands). Close the Serial Monitor to stop live streaming. Finally, disconnect the ESP32 USB and power down the mains feed.
\end{enumerate}

\section{Validation and characterization}

Single-phase immersion cooling has been experimentally shown to reduce cooling energy
consumption compared to conventional approaches \cite{ref5,ref12}. To rigorously
characterize the thermal behavior of the prototype, a structured experimental methodology
was employed, grounded in classical heat transfer theory. The goal was not merely to
demonstrate functionality, but to quantify the system's thermal performance using
well-established metrics that can be compared across dielectric fluids, thermal
loads and future design changes.

\subsection{Theoretical framework and parameter selection}

The thermal behavior of an immersion cooling system can be described using the
analogy between heat transfer and electrical circuits. Just as electrical resistance
governs current flow under a voltage differential, \textit{thermal resistance} governs
heat flow under a temperature differential. This framework --- rooted in Fourier's law
of heat conduction --- motivates the choice of every measured and derived parameter
in this study.

\subsubsection*{Fourier's law and the thermal resistance concept}

The fundamental relation governing steady-state heat conduction is Fourier's law \cite{bergman2011}:

\begin{equation}
    \dot{Q} = -kA\frac{dT}{dx}
    \label{eq:fourier}
\end{equation}

\noindent where $\dot{Q}$ is the rate of heat transfer (W), $k$ is the thermal conductivity
of the medium (W/m\ensuremath{\cdot}K), $A$ is the cross-sectional area (m\textsuperscript{2}), and $dT/dx$ is the
temperature gradient (K/m).

For engineering analysis of multi-interface systems, this is generalized into the
concept of \textit{thermal resistance} $R_{th}$, defined as:

\begin{equation}
    R_{th} = \frac{\Delta T}{\dot{Q}}
    \label{eq:rth}
\end{equation}

\noindent where $\Delta T$ is the temperature difference across an interface (\textdegree{}C or K)
and $\dot{Q}$ is the heat transfer rate through it (W). A lower $R_{th}$ indicates more
effective heat transfer. This metric is dimensionally analogous to Ohm's law
($V = IR$) and enables the decomposition of the total thermal pathway into
distinct, addable resistances.

In this system, two primary thermal interfaces are identified:

\begin{itemize}
    \item \textbf{Heater-to-Fluid interface} ($R_{th,hf}$): Governs how efficiently
    heat is transferred from the heating element surface into the surrounding
    mineral oil. This is dominated by natural convection within the tank.
    \item \textbf{Fluid-to-Ambient interface} ($R_{th,fa}$): Governs how efficiently
    the circulating heated oil rejects its heat to the ambient environment through
    the radiator. This is dominated by forced convection in the external loop.
\end{itemize}

The total system thermal resistance is thus:

\begin{equation}
    R_{th,\text{total}} = R_{th,hf} + R_{th,fa} = \frac{T_{\text{heater}} - T_{\text{ambient}}}{\dot{Q}_{in}}
    \label{eq:rth_total}
\end{equation}

\subsubsection*{Convective heat transfer at the heater surface}

Since the heating element is submerged in a quiescent dielectric fluid with no
forced flow directed at it, heat transfer from the element surface to the bulk
fluid occurs via \textit{natural convection}. This is governed by Newton's Law of Cooling \cite{bergman2011}:

\begin{equation}
    \dot{Q} = h  A_s  (T_s - T_\infty)
    \label{eq:newton}
\end{equation}

\noindent where $h$ is the convective heat transfer coefficient (W/m\textsuperscript{2}\ensuremath{\cdot}K), $A_s$
is the surface area of the heating element (m\textsuperscript{2}), $T_s$ is the heater surface
temperature, and $T_\infty$ is the bulk fluid temperature.

The convective coefficient $h$ for natural convection is typically characterized
using the dimensionless Nusselt number \cite{bergman2011}:

\begin{equation}
    Nu = \frac{h  L_c}{k_f}
    \label{eq:nusselt}
\end{equation}

\noindent where $L_c$ is the characteristic length of the heater (m) and $k_f$ is
the thermal conductivity of the fluid (W/m\ensuremath{\cdot}K). For natural convection, $Nu$ is a
function of the Rayleigh number $Ra$:

\begin{equation}
    Ra = Gr  Pr = \frac{g  \beta  (T_s - T_\infty)  L_c^3}{\nu^2}  \frac{\mu c_p}{k_f}
    \label{eq:rayleigh}
\end{equation}

\noindent where $g$ is gravitational acceleration, $\beta$ is the thermal expansion
coefficient of the fluid, $\nu$ is the kinematic viscosity, $\mu$ is the dynamic
viscosity, and $c_p$ is the specific heat capacity.

Applying Eq.~\ref{eq:newton} to the measured data gives an effective
heat transfer coefficient at the heater surface. Taking the heater's cylindrical
surface (diameter 25.4~mm, length 152.4~mm) as the heat transfer area,
$A_s \approx 0.0122$~m\textsuperscript{2}, the surface heat flux is
$q'' = \dot{Q}_{in}/A_s \approx 41$~kW/m\textsuperscript{2}, and
\begin{equation}
    h_{\text{eff}} = \frac{\dot{Q}_{in}}{A_s\,\Delta T_{hf}} \approx
    \frac{500}{0.0122 \times 89.0} \approx 460~\text{W/m\textsuperscript{2}\ensuremath{\cdot}K}.
    \label{eq:heff}
\end{equation}
Because the element is only partially immersed, the wetted area is smaller than
$A_s$, so this value is a lower bound on the true coefficient. The large
heater-fluid temperature difference ($\Delta T_{hf} = 89.0$~\textdegree{}C) therefore
reflects a high heat flux concentrated on a small surface, transferred to the
oil by buoyancy-driven flow alone. Mineral oil has a significantly higher viscosity than water or fluorocarbon fluids \cite{bergman2011,ref4,ref13}, which limits this
natural convection. Evaluating $Nu$ and $Ra$ from Eqs.~\ref{eq:nusselt}
and~\ref{eq:rayleigh} would require the oil's thermal conductivity, expansion
coefficient and viscosity across the operating temperature range, and is left
for future work.

\subsubsection*{Rationale for measured parameters}

The three sensor locations --- heater surface, bulk fluid, and ambient --- were
deliberately chosen to delineate each stage of the thermal pathway:

\begin{itemize}
    \item $T_{\text{heater}}$ is the \textbf{primary control variable} because it
    directly represents the junction temperature analog of a real server component.
    In data center applications, preventing component overheating is the primary design constraint \cite{ref5,ref12}. Monitoring the heater surface captures
    the worst-case thermal condition in the system.
    \item $T_{\text{fluid}}$ characterizes the \textbf{thermal state of the coolant}
    at the hottest point in the tank (near the warm-fluid outlet). This is the
    driver for the external cooling loop and determines how hard the radiator must
    work.
    \item $T_{\text{ambient}}$ serves as the \textbf{thermodynamic sink reference}.
    Since the radiator rejects heat to the surrounding air, the ambient temperature
    sets the lower bound on achievable fluid temperature and is essential for
    calculating meaningful thermal resistance values.
\end{itemize}

The three derived $\Delta T$ values ($\Delta T_{hf}$, $\Delta T_{fa}$,
$\Delta T_{ha}$) then quantify the temperature drop across each thermal interface.
These are the primary diagnostics for identifying bottlenecks in the cooling chain.

\subsubsection*{Temperature ratio}

The fluid-ambient temperature ratio $\eta_{\text{fluid}}$ used in this study is
defined as the fraction of the total heater-to-ambient temperature difference
that appears across the external cooling loop:

\begin{equation}
    \eta_{\text{fluid}} = \frac{\Delta T_{fa}}{\Delta T_{ha}} =
    \frac{T_{\text{fluid}} - T_{\text{ambient}}}{T_{\text{heater}} - T_{\text{ambient}}}
    \label{eq:efficiency}
\end{equation}

This dimensionless ratio is not an energy efficiency; it captures how much of the system's thermal driving potential is utilized across
the external (fluid-to-ambient) interface. A value close to 1 would indicate
the fluid exits the tank nearly as hot as the heater --- implying the external
loop is the bottleneck. A value close to 0 indicates the fluid barely rises
above ambient --- meaning the in-tank transfer is the bottleneck. Our measured
value of $\eta_{\text{fluid}} = 0.082$ clearly places the internal
heater-to-fluid interface as the dominant resistance, consistent with the
natural convection regime.

\subsection{Experimental methodology}

The prototype was tested under a controlled 500~W constant thermal load, supplied
by the titanium immersion heater. The system was operated with the pump and
radiator fan both active. Data was logged via the ESP32 at 1000~ms intervals using
the DS18B20 sensors (resolution: 0.0625~\textdegree{}C; accuracy: $\pm$0.5~\textdegree{}C from
$-10$~\textdegree{}C to 85~\textdegree{}C and $\pm$2~\textdegree{}C over $-55$~\textdegree{}C to 125~\textdegree{}C \cite{ref14}) until a
steady-state condition was confirmed by the firmware's built-in
steady-state detection flag, which evaluates the rate of change of $T_{\text{heater}}$
over a sliding window. The heater was switched on about 5~minutes after logging
began and ran at full power for about 71~minutes.

Steady-state was declared when:

\begin{equation}
    \left| \frac{dT_{\text{heater}}}{dt} \right| < \epsilon
    \label{eq:steadystate}
\end{equation}

\noindent for a sustained period, where $\epsilon$ is a small threshold defined in
firmware. Because this criterion is based on the heater temperature only, it
identifies a quasi-steady state: the heater temperature had levelled off, but
the bulk oil was still warming slowly (see below).

\subsubsection*{Measurement uncertainty}

The fluid and ambient sensors operate within the $\pm$0.5~\textdegree{}C accuracy band of
the DS18B20, while the heater sensor (above 85~\textdegree{}C) falls in the $\pm$2~\textdegree{}C band
\cite{ref14}. Combining the two sensor errors in the worst case gives
$\Delta T_{fa} = 8.0 \pm 1.0$~\textdegree{}C and $\Delta T_{hf} = 89.0 \pm 2.5$~\textdegree{}C,
and hence $R_{th,fa} = 16.0 \pm 2.0$~m\textdegree{}C/W and
$R_{th,hf} = 178 \pm 5$~m\textdegree{}C/W. The heater input is taken as the element's
500~W rating; the electrical power was not measured, so any deviation of the
supply voltage from nominal adds a further uncertainty to all resistance values.
The heater surface reading also fluctuated by about $\pm$1.6~\textdegree{}C (one standard
deviation) at steady state and reached a peak of 123.8~\textdegree{}C, close to the
sensor's 125~\textdegree{}C rating.

\subsection{Steady-state performance}

Table~\ref{tab:performance_data} presents the steady-state performance
measurements and derived thermal parameters of the immersion cooling system under
a 500~W thermal load. The values are a representative steady-state sample; the
average over the final 10~minutes before the heater was switched off is
practically identical ($T_{\text{heater}} = 118.6 \pm 1.6$~\textdegree{}C,
$T_{\text{fluid}} = 29.4$~\textdegree{}C, $T_{\text{ambient}} = 21.4$~\textdegree{}C,
$R_{th,hf} = 178.4$~m\textdegree{}C/W, $R_{th,fa} = 16.0$~m\textdegree{}C/W,
$\eta_{\text{fluid}} = 0.082$).

\begin{table}[!htbp]
\centering
\caption{Steady-State Performance Data}
\label{tab:performance_data}
\begin{tabular}{p{0.36\linewidth}p{0.14\linewidth}p{0.16\linewidth}p{0.22\linewidth}}
\toprule
\textbf{Parameter} & \textbf{Symbol} & \textbf{Value} & \textbf{Source} \\
\midrule
Heater Input Power & $\dot{Q}_{in}$ & 500 W & Heater rating \\
Heater Temperature & $T_{\text{heater}}$ & 118.3~\textdegree{}C & DS18B20 \\
Fluid Temperature & $T_{\text{fluid}}$ & 29.3~\textdegree{}C & DS18B20 \\
Ambient Temperature & $T_{\text{ambient}}$ & 21.3~\textdegree{}C & DS18B20 \\
$\Delta T$ (Heater--Fluid) & $\Delta T_{hf}$ & 89.0~\textdegree{}C & $T_{\text{heater}} - T_{\text{fluid}}$ \\
$\Delta T$ (Fluid--Ambient) & $\Delta T_{fa}$ & 8.0~\textdegree{}C & $T_{\text{fluid}} - T_{\text{ambient}}$ \\
$\Delta T$ (Heater--Ambient) & $\Delta T_{ha}$ & 97.0~\textdegree{}C & $T_{\text{heater}} - T_{\text{ambient}}$ \\
Temperature ratio (Fluid--Ambient) & $\eta_{\text{fluid}}$ & 0.082 & Calculated via Eq.~\ref{eq:efficiency} \\
Thermal resistance (Heater--Fluid) & $R_{th,hf}$ & 178.0~m\textdegree{}C/W & Calculated via Eq.~\ref{eq:rth} \\
Thermal resistance (Fluid--Ambient) & $R_{th,fa}$ & 16.0~m\textdegree{}C/W & Calculated via Eq.~\ref{eq:rth} \\
Steady-State Indicator & $\textit{steadyState}$ & 1 & ESP32 firmware output \\
\bottomrule
\end{tabular}
\end{table}

The data in Table~\ref{tab:performance_data} tells a clear and consistent
story. Under the full 500~W load, the DS18B20 sensors recorded a heater surface
temperature of 118.3~\textdegree{}C, while the bulk mineral oil stabilized at just 29.3~\textdegree{}C
and the ambient room temperature sat at 21.3~\textdegree{}C. The most striking feature of
these readings is the sheer magnitude of the heater-to-fluid drop: $\Delta T_{hf}
= 89.0$~\textdegree{}C. This means that despite the heater being immersed in oil, the
vast majority of the total system temperature differential --- 89 out of 97~\textdegree{}C, or
roughly 92\% --- is consumed at the single interface between the heater surface and
the surrounding fluid. By contrast, the fluid-to-ambient drop is only 8.0~\textdegree{}C,
meaning the external loop --- the pump, hoses, and radiator together --- handles its
portion of the heat rejection with ease, adding only a modest 8~\textdegree{}C across the
entire circuit from tank outlet to ambient air.

This disparity becomes even more concrete when expressed as thermal resistance.
From Eq.~\ref{eq:rth}, the heater-to-fluid resistance is $R_{th,hf} =
178.0$~m\textdegree{}C/W, while the fluid-to-ambient resistance is $R_{th,fa} =
16.0$~m\textdegree{}C/W --- a ratio of approximately 11:1. In other words, the in-tank
natural convection interface is about eleven times more resistive to heat flow
than the entire external cooling loop. The fluid-ambient temperature ratio of
$\eta_{\text{fluid}} = 0.082$, calculated via Eq.~\ref{eq:efficiency},
reinforces this: only 8.2\% of the total heater-to-ambient temperature
difference appears across the external loop.

Two effects mean that the radiator does not carry the full 500~W, so
$R_{th,fa}$ as defined here is a lower bound on the true radiator-loop
resistance. First, the system was in a quasi-steady rather than fully steady
state: the heater temperature had levelled off, but the bulk oil was still
warming at about 1.8~\textdegree{}C/h when the heater was switched off. For the 35~L
charge ($\approx$28~kg at 808~kg/m\textsuperscript{3} \cite{ref13}) and assuming a typical
mineral-oil specific heat of about 1.9~kJ/kg\ensuremath{\cdot}K \cite{bergman2011}, this corresponds to roughly
25--30~W, or about 5\% of the input, still being stored in the oil. Second,
some heat is lost directly from the glass walls, hoses and pump to the room;
for a wetted tank surface of about 0.5~m\textsuperscript{2} at 8~\textdegree{}C above ambient with
natural-convection coefficients of 5--10~W/m\textsuperscript{2}\ensuremath{\cdot}K \cite{bergman2011}, this is of the order of
20--40~W. Together, these suggest that the radiator rejected roughly
85--90\% of the heater input. None of this affects the main conclusion,
because $R_{th,hf}$ is about eleven times larger.

The calculated thermal resistances are derived directly from Eq.~\ref{eq:rth}:

\begin{align}
    R_{th,hf} &= \frac{\Delta T_{hf}}{\dot{Q}_{in}} = \frac{89.0}{500} = 0.178~\text{\textdegree{}C/W}
    \label{eq:rth_hf_calc} \\[6pt]
    R_{th,fa} &= \frac{\Delta T_{fa}}{\dot{Q}_{in}} = \frac{8.0}{500} = 0.016~\text{\textdegree{}C/W}
    \label{eq:rth_fa_calc} \\[6pt]
    R_{th,\text{total}} &= R_{th,hf} + R_{th,fa} = \frac{97.0}{500} = 0.194~\text{\textdegree{}C/W}
    \label{eq:rth_total_calc}
\end{align}

Together, these values from Table~\ref{tab:performance_data} confirm that
$R_{th,hf}$ accounts for approximately 91.8\% of total system thermal
resistance, while $R_{th,fa}$ contributes only 8.2\%, establishing the
in-tank natural convection as the overwhelmingly dominant bottleneck.

\subsection{Performance summary}

\begin{enumerate}

    \item \textbf{In-tank heat transfer is natural-convection-limited.}
    The heater-fluid temperature difference of 89.0~\textdegree{}C and the associated
    $R_{th,hf}$ of 178.0~m\textdegree{}C/W confirm that heat transport from the heater
    surface to the bulk fluid is governed by natural convection \cite{bergman2011,ref7}. Mineral oil's high kinematic viscosity
    (9.9~cSt at 40~\textdegree{}C) \cite{ref13} significantly suppresses
    buoyancy-driven flow compared to fluorocarbon-based dielectric fluids
    \cite{ref4}, resulting in a lower effective convective coefficient $h$
    and hence a higher interfacial resistance. Earlier comparisons have likewise found mineral oil to remove the least heat among common immersion coolants \cite{ref3}. This could be mitigated in
    future iterations through directional nozzles or mechanical agitation
    targeted at the heater surface.

    \item \textbf{The external cooling loop operates with significant excess capacity.}
    The fluid-to-ambient temperature differential of just 8.0~\textdegree{}C, yielding
    $R_{th,fa} = 16.0$~m\textdegree{}C/W, shows that the radiator and pump can
    reject heat far more readily than the in-tank subsystem can supply it.
    This suggests that the radiator has spare capacity at the current 500~W
    load and could likely support higher thermal loads or operate at reduced
    fan speeds; neither was tested in this study.

    \item \textbf{The heater-fluid interface is the principal thermal bottleneck.}
    The ratio $R_{th,hf} / R_{th,fa} \approx 11.1$ quantifies the disproportion
    between internal and external resistances. In practical terms, improving the
    in-tank convection regime --- for example, by implementing jet impingement,
    directional flow channels, or switching to a lower-viscosity dielectric fluid
    such as FC-40 \cite{ref4} --- would yield far greater improvements in overall
    system performance than upgrading the external radiator or pump.

    \item \textbf{The system reaches stable quasi-steady operation.}
    The prototype ran for about 71~minutes under a continuous 500~W load, with
    the heater temperature levelled off and the steady-state flag set by the
    ESP32 firmware. The fluid temperature reached 29.3~\textdegree{}C --- only 8~\textdegree{}C above
    ambient and still rising slowly at about 1.8~\textdegree{}C/h --- indicating that the coolant loop remained well within its safe operating
    range and is suitable for controlled repeatability studies.

\end{enumerate}

The performance statistics across the full experimental run are shown in
Fig.~\ref{fig:performance} (see \hyperref[sec:appendixB]{Appendix~B}
for full dataset).

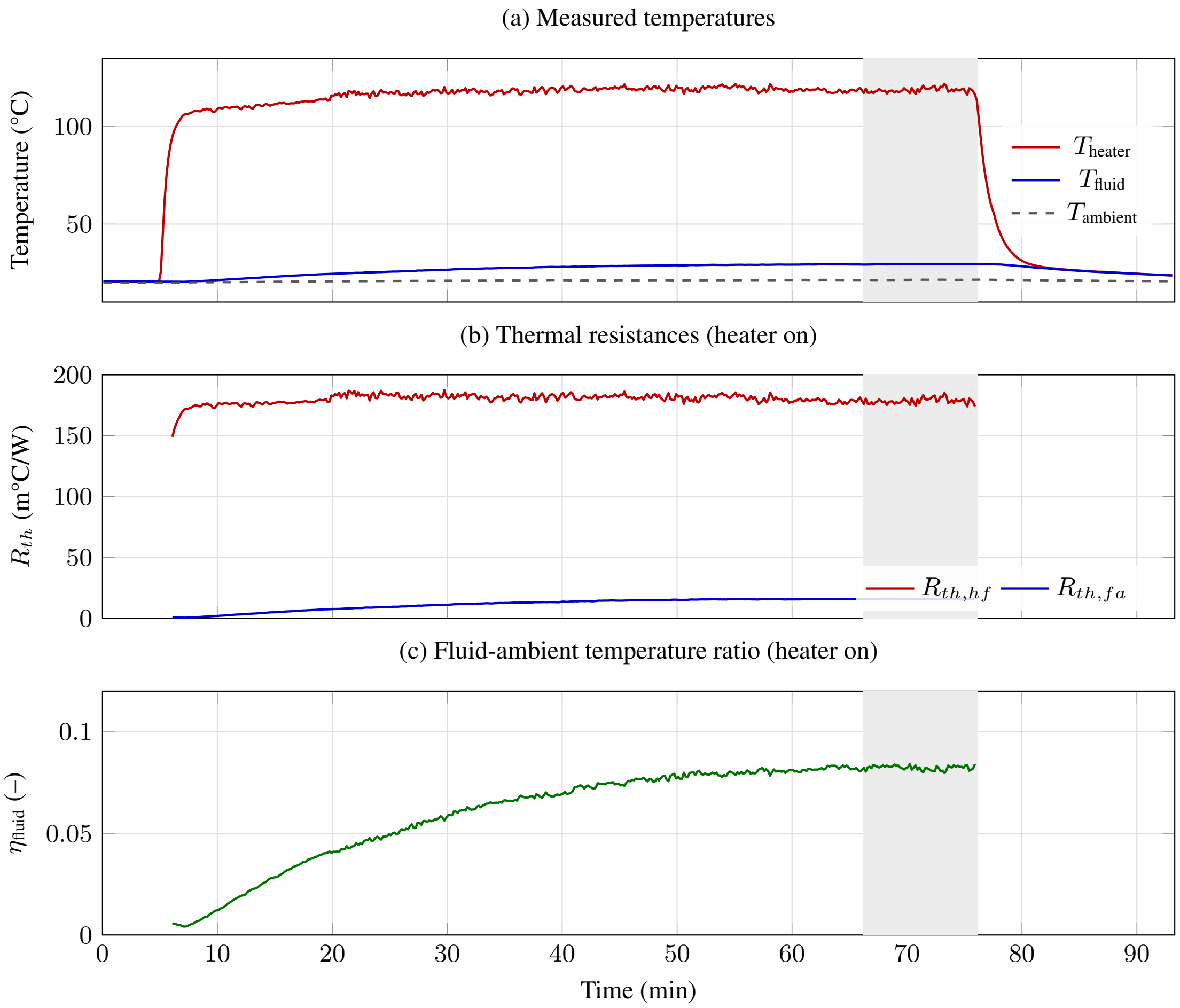
\begin{figure}[!htbp]
    \centering
    \begin{tikzpicture}
    \begin{groupplot}[
        group style={group size=1 by 3, vertical sep=0.9cm, xlabels at=edge bottom, xticklabels at=edge bottom},
        width=0.9\linewidth, height=4.6cm,
        xmin=0, xmax=93.3, xlabel={Time (min)},
        grid=major, grid style={gray!25},
        tick label style={font=\small}, label style={font=\small},
        legend style={font=\footnotesize, draw=none, fill=white, fill opacity=0.85, text opacity=1},
        unbounded coords=jump, no markers,
        every axis plot/.append style={line width=0.8pt, line join=round},
    ]
    \nextgroupplot[ylabel={Temperature (\textdegree{}C)}, ymin=10, ymax=135,
        legend style={at={(0.98,0.5)}, anchor=east}, legend columns=1,
        title={(a) Measured temperatures}, title style={font=\small}]
    \fill[gray!15] (axis cs:66.15,10) rectangle (axis cs:76.15,135);
    \addplot[red!75!black] table[x=t, y=Th] {images/performance.dat};
    \addlegendentry{$T_{\text{heater}}$}
    \addplot[blue!80!black] table[x=t, y=Tf] {images/performance.dat};
    \addlegendentry{$T_{\text{fluid}}$}
    \addplot[gray!70!black, dashed] table[x=t, y=Ta] {images/performance.dat};
    \addlegendentry{$T_{\text{ambient}}$}
    \nextgroupplot[ylabel={$R_{th}$ (m\textdegree{}C/W)}, ymin=0, ymax=200,
        legend pos=south east, legend columns=2,
        title={(b) Thermal resistances (heater on)}, title style={font=\small}]
    \fill[gray!15] (axis cs:66.15,0) rectangle (axis cs:76.15,200);
    \addplot[red!75!black] table[x=t, y=Rhf] {images/performance.dat};
    \addlegendentry{$R_{th,hf}$}
    \addplot[blue!80!black] table[x=t, y=Rfa] {images/performance.dat};
    \addlegendentry{$R_{th,fa}$}
    \nextgroupplot[ylabel={$\eta_{\text{fluid}}$ (--)}, ymin=0, ymax=0.12,
        yticklabel style={/pgf/number format/fixed, /pgf/number format/precision=2},
        title={(c) Fluid-ambient temperature ratio (heater on)}, title style={font=\small}]
    \fill[gray!15] (axis cs:66.15,0) rectangle (axis cs:76.15,0.12);
    \addplot[green!45!black] table[x=t, y=eta] {images/performance.dat};
    \end{groupplot}
    \end{tikzpicture}
    \caption{Time series of the full experimental run, shown as 10-second
    averages of the 1~Hz data. The heater was switched on at about 5~min and off
    at about 76~min. (a) Heater, fluid and ambient temperatures. (b) Thermal
    resistances $R_{th,hf}$ and $R_{th,fa}$, computed with
    $\dot{Q}_{in} = 500$~W. (c) Fluid-ambient temperature ratio
    $\eta_{\text{fluid}}$. Panels (b) and (c) are shown only while the heater is
    on, because they are undefined without a heat input. The shaded band marks
    the final 10~minutes of heating used for the steady-state averages.}
    \label{fig:performance}
\end{figure}

\subsection{Capabilities}

\begin{enumerate}
    \item The system supports stable and continuous operation under a 500~W constant
    thermal load, demonstrating the viability of single-phase mineral oil immersion
    cooling at moderate power densities.
    \item A stable quasi-steady state is reached with active circulation and
    air-cooled heat rejection, showing closed-loop stability under the implemented
    threshold-based control strategy.
    \item Real-time monitoring of $T_{\text{heater}}$, $T_{\text{fluid}}$, and
    $T_{\text{ambient}}$ via DS18B20 sensors enables live computation and logging
    of $R_{th}$, $\Delta T$, and $\eta_{\text{fluid}}$, making the platform
    suitable as an experimental testbed.
    \item The modular design permits straightforward substitution of components ---
    including alternative dielectric fluids, higher-capacity pumps, larger radiators,
    or different tank geometries --- enabling parametric studies without redesigning
    the system.
    \item The complete system is assembled for INR~35,318 (approximately
    USD~380), well within the INR~50,000 project budget and a small fraction of
    the cost of commercial immersion cooling systems.
\end{enumerate}

\subsection{Limitations}

\begin{enumerate}
    \item \textbf{Simulated thermal load.} The 500~W titanium heating element
    approximates uniform volumetric heat generation but does not replicate the
    spatially non-uniform heat flux profile of real server hardware, where hotspots
    occur at specific components such as the CPU, GPU, and VRMs \cite{ref5}.
    This may lead to an underestimation of local peak temperatures in a real
    deployment.

    \item \textbf{Natural convection dominance.} The absence of forced flow within
    the tank --- no jet impingement, no internal nozzles, no baffle-directed flow ---
    means heat transfer from the heater surface to the fluid is entirely driven by
    buoyancy effects. This is the primary reason $R_{th,hf}$ is so large relative
    to $R_{th,fa}$, and it is a fundamental constraint of the current design.
    Alternative low-viscosity dielectric fluids such as FC-40 \cite{ref4} could
    partially alleviate this by improving buoyancy-driven circulation.

    \item \textbf{No internal flow shaping.} Without directional nozzles or flow
    channels, the circulation provided by the pump acts only on the bulk fluid and
    does not enhance localized convection at the heater surface. Future designs
    should incorporate internal manifolds to direct cooled fluid toward heat-generating surfaces; in a forced-circulation immersion system, raising the coolant velocity across the heat sink has been shown to lower CPU temperature and reduce flow stagnation around it \cite{ref3}.

    \item \textbf{Manual fluid management.} The system lacks a drain port, overflow
    valve, or filtration loop. Oil filling and draining must be performed manually,
    increasing the risk of spills, contamination of the dielectric fluid over time,
    and making long-term maintenance cumbersome.

    \item \textbf{Single test condition and indirect heat balance.} Results are
    from one 500~W run at one pump and fan setting. The heater power is taken
    from its rating rather than measured, and the oil-side flow rate and
    radiator heat duty were not measured, so the radiator-loop resistance is
    inferred rather than obtained from a direct energy balance. Adding a power
    meter, a flow meter and inlet/outlet temperature sensors on the radiator
    would close this gap.

    \item \textbf{Sensor range.} The heater-surface DS18B20 is rated to 125~\textdegree{}C
    and its accuracy degrades to $\pm$2~\textdegree{}C above 85~\textdegree{}C \cite{ref14}. The peak
    reading of 123.8~\textdegree{}C leaves little margin, so higher thermal loads would need
    a thermocouple or RTD at the heater surface.

    \item \textbf{Temperature ceiling imposed by glass construction.} The glass tank,
    while providing excellent visibility for experimental purposes, has limited
    mechanical and thermal tolerance compared to stainless steel
    enclosures. Thermal cycling and elevated operating temperatures risk stress
    fractures or sealant degradation, restricting experimental operation to moderate
    temperature regimes.
\end{enumerate}

\section*{Declaration of generative AI and AI-assisted technologies in the manuscript preparation process}

During the preparation of this work, the author(s) used ChatGPT, Google Gemini, and Claude AI in order to improve language clarity, locate citations within the text, generate LaTeX formatting code, and assist with technical documentation structure. After using these tools, the author(s) reviewed and edited the content as needed and take(s) full responsibility for the content of the manuscript.

\section*{Ethics statements}

This study did not involve human subjects, personal data, or biological specimens. No animal experiments were performed, and therefore, animal-research guidelines are not applicable to this work. This whole project only involved mechanical and electronic hardware.

\section*{CRediT author statement}

\textbf{Pramukto Mandal}: Software, Investigation, Data Curation. \textbf{Bhavya Jain}: Methodology, Visualization. \textbf{M.U. Kusheetha}: Writing - Original Draft, Conceptualization. \textbf{Nikhil Meena}: Resources, Software. \textbf{Ravindra Kumar}: Conceptualization, Investigation. \textbf{M.P. Sabarish}: Conceptualization, Visualization, Writing - Original Draft. \textbf{Amit Verma}: Supervision, Project Administration. \textbf{Gaurav Bhutani}: Supervision, Writing - Review \& Editing.

\section*{Acknowledgements}

The authors would like to thank IIT Mandi for providing access to the required laboratory facilities and equipment essential for the successful completion of this project. This work was developed as part of the course IC-202P, and we express our sincere gratitude to the course instructor, Dr. Gajendra Singh, for his guidance throughout the development of this project.

\textbf{Funding}: This work was supported by the Indian Institute of Technology Mandi, through an internal student project fund of INR 50,000.

\bibliographystyle{unsrtnat}
\bibliography{references}

\section*{Appendix A: ESP32 control firmware}
\label{sec:appendixA}

The complete firmware (sensor acquisition, temperature-based relay control, manual override and safety protection) is provided as open source code in the file \texttt{fin.ino} of the project repository:

\noindent\url{https://github.com/koc003/Design-and-Experimental-Evaluation-of-a-Low-Cost-Immersion-Cooling-System-ESP32}

\section*{Appendix B: performance data}
\label{sec:appendixB}

The complete time-series dataset is the file \texttt{data.csv} in the same repository. It has no header row and one row per second. Its ten columns are, in order: $T_{\text{heater}}$, $T_{\text{fluid}}$, $T_{\text{ambient}}$ (all in \textdegree{}C), $\Delta T_{hf}$, $\Delta T_{fa}$, $\Delta T_{ha}$ (\textdegree{}C), $\eta_{\text{fluid}}$, $R_{th,hf} \times 1000$ and $R_{th,fa} \times 1000$ (m\textdegree{}C/W, computed with $\dot{Q}_{in} = 500$~W), and the firmware steady-state flag (0 or 1). The resistance and ratio columns are only meaningful while the heater is on (approximately rows 300 to 4570).

\end{document}